\documentclass[aps,
prd,
twocolumn,
superscriptaddress,
showpacs,
preprintnumbers,
nofootinbib,
notitlepage]{revtex4-1}

\usepackage{amsmath,amssymb}
\usepackage{hyperref}
\usepackage{subfigure,xspace}
\usepackage{color,graphicx}
\usepackage{braket,bbm,bm}
\usepackage{multirow}
\usepackage{enumerate}
\usepackage{xspace}
\usepackage{wrapfig}
\usepackage{comment}
\usepackage{feynmp-auto}
\usepackage[english]{babel}
\usepackage[export]{adjustbox}
\usepackage[utf8]{inputenc}

\renewcommand{\k}{\bm{k}}
\newcommand{\p}{\bm{p}}
\newcommand{\q}{\bm{q}}

\renewcommand{\P}{\bm{P}}

\newcommand{\0}{\bm{0}}
\newcommand{\1}{\bm{1}}
\newcommand{\2}{\bm{2}}
\newcommand{\3}{\bm{3}}

\newcommand{\n}{\bm{n}}
\newcommand{\m}{\bm{m}}

\newcommand{\one}{\mathbbm{1}}

\newcommand{\Ac}{\mathcal{A}}
\newcommand{\wAc}{\widetilde{\mathcal{A}}}
\newcommand{\Bc}{\mathcal{B}}
\newcommand{\Cc}{\mathcal{C}}
\newcommand{\Dc}{\mathcal{D}}

\newcommand{\Fc}{\mathcal{F}}
\newcommand{\Gc}{\mathcal{G}}
\newcommand{\Hc}{\mathcal{H}}
\newcommand{\Ic}{\mathcal{I}}
\newcommand{\Jc}{\mathcal{J}}
\newcommand{\Kc}{\mathcal{K}}
\newcommand{\Lc}{\mathcal{L}}
\newcommand{\Mc}{\mathcal{M}}

\newcommand{\Oc}{\mathcal{O}}

\newcommand{\Rc}{\mathcal{R}}

\newcommand{\Tc}{\mathcal{T}}

\DeclareMathOperator{\re}{Re}
\DeclareMathOperator{\im}{Im}

\newcommand{\bs}[1]{\boldsymbol{ #1 }}

\newcommand{\bh}[1]{\bf{\hat{ #1 }}}

\renewcommand{\min}{{\text{min}}}
\renewcommand{\bra}{ \langle }
\renewcommand{\ket}{ \rangle }
\newcommand{\nln}{\\ \nonumber}

\newcommand{\bcol}{\left[ \begin{array}{c}}
\newcommand{\ecol}{\end{array} \right]}
\newcommand{\beq}{\begin{eqnarray}}
\newcommand{\eeq}{\end{eqnarray}}

\newcommand{\kev}{\ensuremath{{\mathrm{\,ke\kern -0.1em V}}}\xspace}
\newcommand{\mev}{\ensuremath{{\mathrm{\,Me\kern -0.1em V}}}\xspace}
\newcommand{\gev}{\ensuremath{{\mathrm{\,Ge\kern -0.1em V}}}\xspace}
\newcommand{\tev}{\ensuremath{{\mathrm{\,Te\kern -0.1em V}}}\xspace}

\usepackage{mfirstuc} 
\newcommand{\addReviewer}[2]{
  \expandafter\newcommand\csname #1\endcsname[1]{{\bf \color{#2} \capitalisewords{#1}:\,##1}}
  \expandafter\newcommand\csname #1cor\endcsname[2]{{\color{#2} \capitalisewords{#1}:\,\st{##1}{\bf ##2}}}
  \expandafter\newcommand\csname #1color\endcsname{#2}
}

\usepackage{soul,color}
\definecolor{chromeyellow}{rgb}{1.0, 0.65, 0.0}
\definecolor{DodgeBlue}{rgb}{0.118, 0.565,1.000}
\definecolor{asparagus}{rgb}{0.53, 0.66, 0.42}
\definecolor{cardinal}{rgb}{0.77, 0.12, 0.23}
\definecolor{cadmiumgreen}{rgb}{0.0, 0.42, 0.24}
\definecolor{applegreen}{rgb}{0.55, 0.71, 0.0}

\addReviewer{wrong}{red}
\addReviewer{good}{green}
\addReviewer{sebastian}{cardinal}

\hypersetup{ 
pdfnewwindow=true,
colorlinks=true,    
linkcolor=blue,
citecolor=blue,
filecolor=blue,
urlcolor=blue
} 

\begin{document}

\preprint{JLAB-THY-20-3274}

\newcommand{\ceem}{Center for Exploration of Energy and Matter, Indiana University, Bloomington, IN 47403, USA}
\newcommand{\indiana}{Physics Department, Indiana University, Bloomington, IN 47405,  USA}
\newcommand{\jlab}{Theory Center, Thomas Jefferson National Accelerator Facility, Newport News, VA 23606,  USA}


\title{Bound states in the B-matrix formalism for the three-body scattering}


\author{Sebastian~M.~Dawid}
\email[email: ]{sdawid@iu.edu}
\affiliation{\indiana}
\affiliation{\ceem}

\author{Adam~P.~Szczepaniak}
\email[email: ]{aszczepa@iu.edu}
\affiliation{\indiana}
\affiliation{\ceem}
\affiliation{\jlab}


\begin{abstract}
We consider a model of relativistic three-body scattering with a bound state in the two-body sub-channel. We show that the na\"ive $K$-matrix type parametrization, here referred to as the $B$-matrix, has nonphysical singularities near the physical region. We show how to eliminate such singularities by using dispersion relations and also show how to reproduce unitarity relations by taking into account all relevant open channels.    
\end{abstract}

\date{\today}
\maketitle

\section{Introduction}

Strong interactions between quarks and gluons give rise to a rich spectrum of resonances, many of which decay into three or more hadrons. To understand this aspect of the QCD phenomenology it is necessary to construct analytic reaction amplitudes. This is because the extraction of resonance parameters from experimental data~\citep{Bookwalter:2011, Adolph:2014, Adolph:2015b, Ghoul:2015, Adolph:2015a, Esposito:2016, AlGhoul:2017, Lebed:2016, Olsen:2017, Akhunzyanov:2018} and lattice QCD (LQCD) simulations of hadron scattering~\citep{Wilson:2014, Lang:2015, Moir:2016, Briceno:2016, Dudek:2016, Andersen:2017, Briceno:2017a, Briceno:2017c, Mai:2018, Guo:2018l, Andersen:2018, Brett:2018, Woss:2018, Woss:2019, Culver:2019, Horz:2019, Hansen:2019l} requires a continuation of the amplitudes in energies and momenta outside of the physical region. It is particularly important for the identification of new states and the determination of their nature, e.g. in the context of the quark model classification. For example, the $a_1(1420)$ decaying into three pions~\citep{Ketzer:2015, Basdevant:2015} may result from a kinematical reflection rather than being a genuine resonance, while the  $\chi_{c1}(3872)$ (also known as $X(3872)$) decaying to $D^0\bar{D}^0\pi$ and $J/\psi \pi\pi$~\citep{Choi:2003, Kang:2016j} may be a di-meson molecule instead of a compact quark bound state.

Two classes of relativistic three-body approaches are currently being pursued, especially in connection with future analyses of LQCD simulations. One is based on linear equations, which in effect sum up particle exchange interactions. These are often motivated by a generic relativistic effective field theory (EFT)~\citep{Hansen:2014, Hansen:2015, Hansen:2016, Briceno:2017, Mai:2017a, Briceno:2018a, Briceno:2019muc, Blanton:2019a, Hansen:2020, Blanton:2020a, Blanton:2020b}. The other follow the $S$-matrix philosophy by applying unitarity constraints~\citep{Mai:2017b, Jackura:2018, Mikhasenko:2019, Jackura:2019} to determine imaginary parts of the on-shell amplitudes. The real parts are either parametrized using the three-body analog of the two-body $K$ matrix referred to as the $B$ matrix, or derived from analyticity. In the following, we refer to the two approaches as the EFT and the $B$-matrix, respectively. They have recently been applied to the study of various three-body phenomena~\citep{Briceno:2018, Romero-Lopez:2018, Romero-Lopez:2019, Blanton:2019, Mai:2019, Sadasivan:2020, Hansen:2020otl}.

Both formalisms lead to a representation of the three-body amplitude that has a form of an integral equation. The kernel in this equation may contain both the long-range, one-particle-exchange amplitude (OPE) and short-range interactions. In practice, a description of the off-shell part of the OPE kernel is the main difference between the EFT and $B$-matrix approaches. In the EFT framework, the OPE kernel contains both real and virtual components while in the $B$-matrix approach, where all amplitudes are on-shell, the former is not explicit but can be included in the short-range part of the $B$-matrix kernel. As demonstrated in Refs.~\citep{Jackura:2019, Blanton:2020b}, in both the infinite and finite volume, the two approaches are equivalent, albeit related by a set of complicated integral relations. In practice, however, since the off-shell effects or left-hand cuts are often parametrized in each analysis independently, results from the two frameworks may be different.

Our interest in this paper is primarily in assessing the suitability of the $B$-matrix formalism in studying the formation of three-body bound states. Therefore we consider a simplified model of the three-body scattering, in which the long-range one-particle exchanges are neglected, and only short-range contact interactions are included. Furthermore, it is assumed that a bound-state can develop in the two-body sub-channel.

The analysis leads us to the conclusion that ``simple" short-range, contact interaction kernels in the $B$-matrix formalism result in the three-body amplitudes with nonphysical analytic properties. In particular, in the presence of singularities near the two-particle threshold, spurious singularities may appear arbitrarily close to the three-body threshold and hinder the formation of genuine three-body singularities: poles or virtual states. Moreover, in the $B$-matrix formalism, one cannot obtain the bound-state--particle amplitude from the three-body amplitude. One would do this by setting the incoming and outgoing isobar energies equal to the two-body bound-state energy~\citep{Bedaque:1997, Bedaque:1998, Jackura:2020bsk}, which is, however, outside the physical region of the $\3 \to \3$ amplitude and thus affected by the spurious left-hand cuts. 

We show that to obtain the correct results in the $B$-matrix approach, it is necessary to explicitly include the channel representing bound-state--particle scattering. Furthermore, since the $B$-matrix formalism builds upon on-shell, unitary amplitudes, to remove the undesired left-hand singularities it is necessary to use  dispersion relations. It would be interesting to see what analytic behavior emerges in other, e.g. EFT approaches when using such interactions.

The paper is organized as follows. In Sec.~\ref{sec:b-matrix}, following Refs.~\citep{Jackura:2018, Jackura:2019}, the $B$-matrix framework for $\3 \to \3$ scattering is reviewed. We compare the $B$-matrix ladder equation with the EFT equivalents and pinpoint the key differences between the two. In Sec.~\ref{sec:multi-channel}, we introduce the multi-particle generalization of the $B$-matrix representation, based on the multi-channel unitarity of the $S$-matrix and give the formal solutions for the amplitudes. In Sec.~\ref{sec:contact}, we present the short-range interaction model for $S$-wave scattering and discuss the dispersive representation, which removes the nonphysical singularities. Conclusions and outlook are summarized in Sec.~\ref{sec:conclusions}. The paper contains four appendices. In App.~\ref{app:A}, the three-body kinematics and conventions are explained. In App.~\ref{app:B}, we show the relation between the non-relativistic EFT (NREFT) scattering amplitude of Ref. \citep{Bedaque:1998} and the non-relativistic approximation of the three-body $B$-matrix ladder equation. In App.~\ref{app:C}, the multi-channel unitarity relations are presented, together with the proof that they are satisfied by the generalized $B$-matrix representation. Finally, in App. \ref{app:D} attached are additional figures to illustrate the discussion of the analytic structures.

\section{The B-matrix parametrization}
\label{sec:b-matrix}

\begin{figure}[t!]
    \centering
    \includegraphics[ width=0.95\columnwidth, trim= 4 4 4 4,clip]{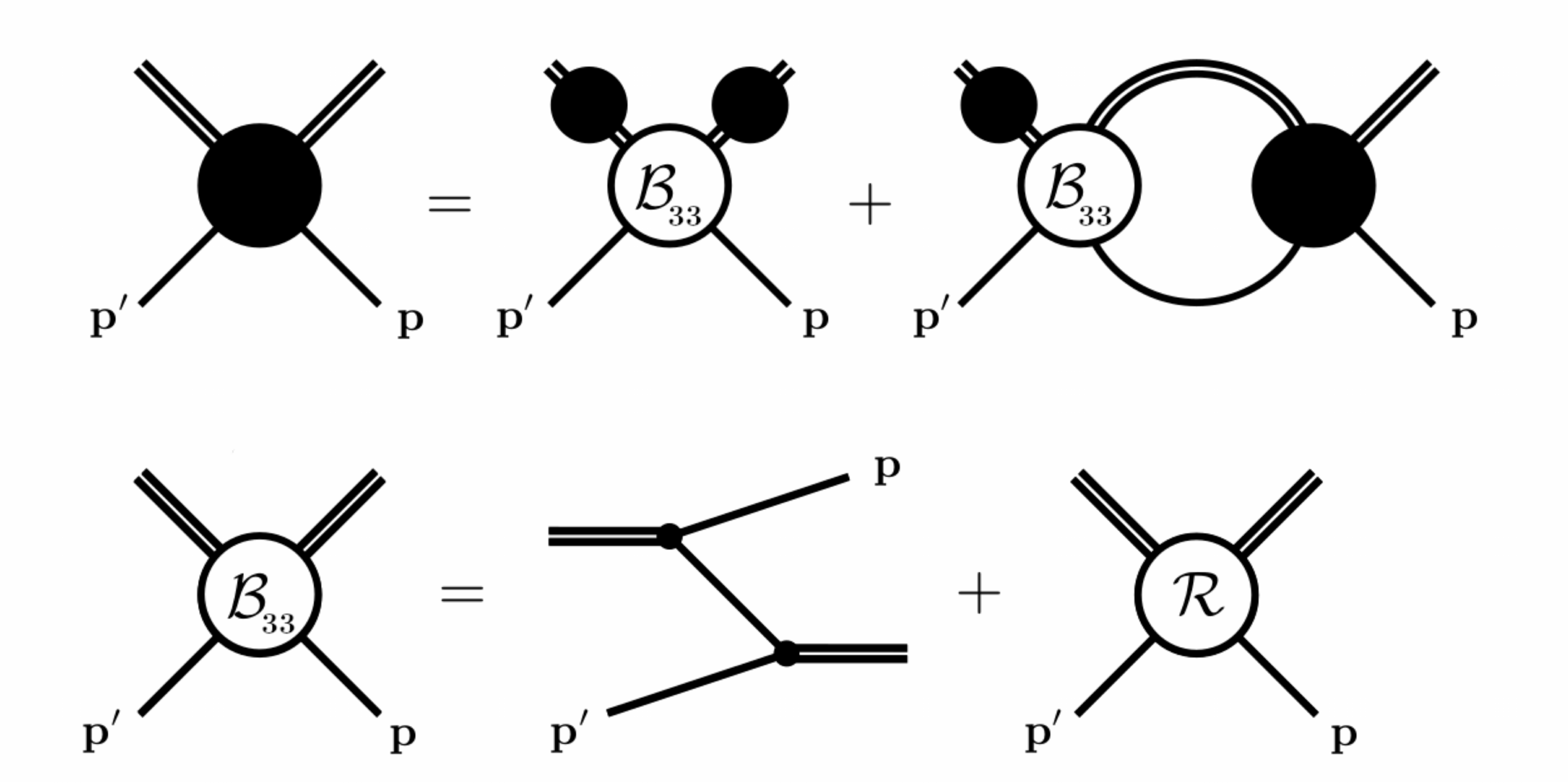}
    \put(-245,90){\colorbox{white}{(a)}}
    \put(-245,25){\colorbox{white}{(b)}}
    \caption{
    Diagrammatic representation of (a) $\Ac_{33}$, as given by Eq.~\eqref{eq:b-matrix-param}, and (b) the $B$-matrix kernel of Eq.~\eqref{eq:B-matrix-decomp}. As explained in App.~\ref{app:A}, a single external line represents a spectator, while a double external line---an isobar. A solid circle with both external isobars and spectators is the three-body connected amplitude $\Ac_{33,\p'\p}$, and a solid circle only with external isobars is the two-body amplitude $\Fc_{\p}$.}
    \label{fig:b-matrix}
    \end{figure}

We start with a brief overview of the $B$-matrix formalism and the $\3 \to \3$ scattering introduced in Refs.~\citep{Jackura:2018, Mikhasenko:2019, Jackura:2019}. A summary of notation and normalization conventions, which are adopted from Ref.~\citep{Jackura:2019}, is given in App.~\ref{app:A}. In particular, the three-body amplitude $\Mc_{33}$, for spin-0 particles, and its unsymmetrized partial-wave projected version $\Mc_{33,\p'\p}$, are defined in Eqs.~\eqref{eq:Smatrix} and~\eqref{eq:SymmAmp}, respectively. The $B$-matrix parametrization for the connected part $\Ac_{33,\p'\p}$ of the amplitude $\Mc_{33,\p'\p}$ is given by the matrix-integral linear equation
    \beq
    \label{eq:b-matrix-param}
    \Ac_{33,\p' \p} = \Fc_{\p'} \, \Bc_{33,\p' \p} \, \Fc_{\p} + \int_{\q} \Fc_{\p'} \, \Bc_{33,\p' \q} \, \Ac_{33,\q \p} \, ,
    \eeq
as illustrated in Fig.~\ref{fig:b-matrix}. The amplitude $\Ac_{33,\p' \p}$ and the kernel $\Bc_{33,\p'\p}$ are matrices in the space labeled by the spin $(\ell, m_\ell)$ of the isobar. The isobar is defined through the $\2 \to \2$ partial wave amplitude $\Fc_{\p}$. The product of $\Fc_{\p}$ and the momentum conserving delta function for the spectator defines the disconnected part of $\Mc_{33,\p'\p}$, as in Eq.~\eqref{eq:dis-con-decomp}. The $B$-matrix kernel is written as a sum of two terms,
    \beq
    \label{eq:B-matrix-decomp}
    \Bc_{33,\p'\p} = \Gc_{\p'\p} + \Rc_{\p'\p} \, ,
    \eeq
where the matrix $\Gc_{\p'\p}$ represents the long-range interaction due to one-particle exchange (OPE) between the isobar and spectator and $\Rc_{\p' \p}$ is a real matrix that absorbs all short-range interactions.

Analytic structure of the $S$-wave OPE, as a function of the total invariant mass squared $s$, for fixed subchannel invariant masses $\sigma_{\p'}$, $\sigma_{\p}$, was explored in Ref.~\citep{Jackura:2018}. A single iteration of the kernel $\Bc_{33,\p'\p}$ generates so-called: bubble ($\Rc\times \Rc$), triangle ($\Rc \times \Gc$) and box ($\Gc \times \Gc$) diagrams. Since these are determined by direct channel unitarity only, they do not have the analytical structure of covariant Feynman amplitudes. This results in spurious left-hand cuts, and a dispersion prescription was proposed as a way to remove them. It was presented for the triangle diagram, and, as shown in Sec.~\ref{sec:contact}, a similar situation is found in our model, which effectively sums up a series of bubble diagrams.

    \begin{figure*}[ht!]
    \centering
    \includegraphics[ width=0.8\textwidth]
    {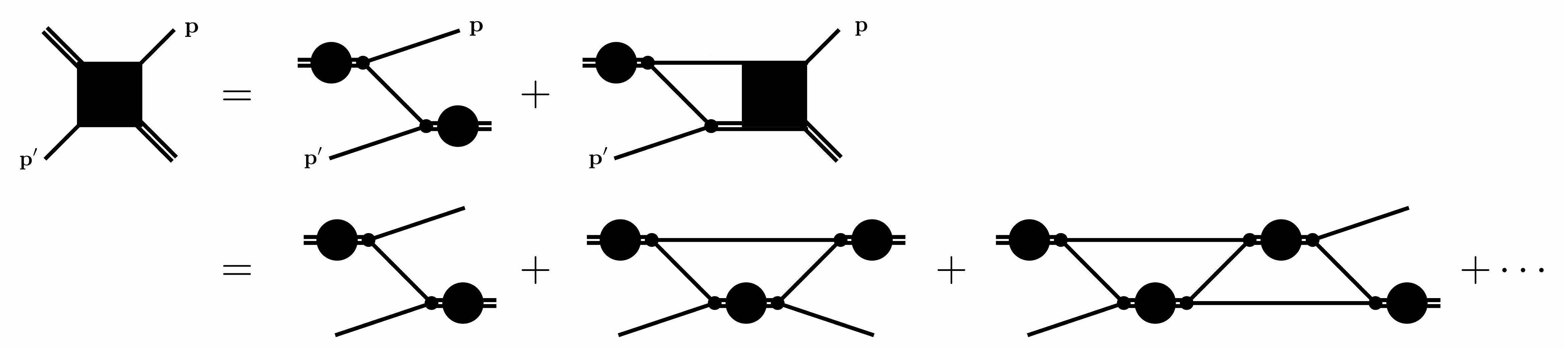}
    \caption{Diagrammatic representation of the ladder Eq.~\eqref{fig:ladder}. Here the black box represents $\Dc_{\p'\p}$.}
    \label{fig:ladder}
    \end{figure*}

Equation~\eqref{eq:b-matrix-param} can be considered in the so-called ladder approximation, in which only the OPE amplitude is included in the $\Bc_{33,\p'\p}$ kernel, leading to the solution driven exclusively by the exchanges between $\2\to\2$ sub-processes. Namely, assuming $\Rc_{\p'\p}=0$ and defining the amplitude given by such an equation as $\Dc_{\p'\p}$, one obtains
    \beq
    \label{eq:ladder-1}
    \Dc_{\p' \p} = \Fc_{\p'} \, \Gc_{\p' \p} \, \Fc_{\p} + \int_{\q} \Fc_{\p'} \, \Gc_{\p' \q} \, \Dc_{\q \p} \, .
    \eeq
This is what is referred to as the ladder amplitude. Figure~\ref{fig:ladder} shows a diagrammatic representation of the ladder series solution. The connected amplitude $\Ac_{33,\p'\p}$ can be rewritten as \citep{Mikhasenko:2019, Jackura:2019}
    \beq
    \label{eq:ladder-contact-decomp}
    \Ac_{33,\p' \p} = \Dc_{\p' \p} + \int_{\q} \int_{\q'} \widetilde{\Lc}_{\p'\q'} \, \widetilde{\Tc}_{\q' \q} \, \widetilde{\Lc}_{\q \p} \, ,
    \eeq
where the amplitude $\widetilde{\Tc}_{\p'\p}$ is given by the integral equation
    \beq
    \label{eq:divergence-free-amplitude}
    \widetilde{\Tc}_{\p' \p} = \Rc_{\p' \p} + \int_{\q} \int_{\q'} \Rc_{\p'\q'} \, \widetilde{\Lc}_{\q' \q} \, \widetilde{\Tc}_{\q \p} \, ,
    \eeq
with $\widetilde{\Lc}_{\p'\p} = \Fc_{\p'\p} \, \delta_{\p'\p} + \Dc_{\p'\p}$. In this form, Eq.~\eqref{eq:b-matrix-param} is explicitly split into terms that are generated individually by either the OPE or the $\Rc$ matrix. Since, as discussed below, different formalisms of the three-body scattering differ by $\Rc$, while the OPE amplitude is universal, Eq.~\eqref{eq:ladder-1} needs to be solved only once for a given $\Fc$, and there are ongoing efforts to calculate the ladder amplitude in the presence of the bound states~\citep{Jackura:2020bsk}. Once the ladder amplitude is known, one can include the short-range effects via the solution of  Eq.~\eqref{eq:divergence-free-amplitude}. The distinction between the long-range particle exchange and genuine short-range interactions due to virtual exchanges becomes important in addressing the nature of shallow bound states or resonances that are produced close to the opening of a three-particle channel. An example of such system, the $\chi_{c1}(3872)$ mentioned earlier, has a mass only about $7\mbox{ MeV}$ and $0.4\mbox{ MeV}$ away from the $D^0 \bar{D}^0 \pi^0$ and $D^0 \bar{D}^{*0}$ thresholds, respectively. It was hypothesized to be a hadronic molecule of $\bar{D}^{*0}$ and $D^0$ bound via a pion exchange \citep{Thomas:2008, Braaten:2010, Baru:2011, Kalashnikova:2012, Guo:2017}. Such a model can be conveniently addressed by the $B$-matrix formalism, with its clear distinction between the short-range and long-range amplitudes. Namely, if the ladder amplitude alone was sufficient in generating the resonance, it would strongly suggest the molecular interpretation.

Equation~\eqref{eq:b-matrix-param} is a general representation of the on-shell amplitude based on the principle of unitarity, therefore it is expected that any formalism of the $\3 \to \3$ scattering can be rewritten in this form~\citep{Jackura:2019, Blanton:2020b}. The difference between various formalisms lies mainly in the definition of the short-range interaction kernel $\Rc_{\p' \p}$. For example, the divergence-free $K$ matrix of the relativistic EFT formalism of Ref.~\citep{Hansen:2015} can be transformed into the $R$ matrix via a complicated integral formula, (see Eq.~(31) of Ref.~\citep{Jackura:2019}). The ladder equations in both frameworks look formally identical, since they do not involve the short-range interactions. However, there is a significant difference coming from the way the integration range over the intermediate particles momenta in both formalism is defined. In general,
    \beq
    \label{eq:integration}
    \int_{\q} &\equiv& \! \int \! \frac{d \Omega_{\q}}{4 \pi}\!\!\! \int\limits_0^{q_{\text{max}}} \!\! \frac{d q \ q^2}{2\pi^2 \omega_{q} } \! = \!
    \int \! \frac{d \Omega_{\q}}{4 \pi} \!\!\!\! \int\limits_{\sigma_{\text{min}}}^{(\sqrt{s}-m)^2} \!\!\! \frac{d \sigma_{\q}}{2 \pi} \, \tau(s, \sigma_{\q}) \, .
    \eeq
Here $\tau(s, \sigma_{\q}) = \lambda^{1/2}(s, \sigma_{\q}, m^2) / 8 \pi s$ is the three-body phase space factor. The relation between $|\q|$, which is the magnitude of the spectator momentum in the three particle rest frame, and the invariant mass of the isobar, $\sigma_{\q}$ is given by
    \beq
    \sqrt{s} = \sqrt{m^2 + \q^2} + \sqrt{\sigma_{\q} + \q^2} \, ,
    \eeq 
or
    \beq
    |\q| = \frac{\lambda^{1/2}(\sigma_{\q}, s , m^2)}{2\sqrt{s}} \, .
    \eeq
At fixed $s$, increasing the UV momentum cut-off $q_{\text{max}}$, corresponds to the decreasing of the lower limit $\sigma_{\text{min}}$ in the integral over $\sigma_{\q}$. The $B$-matrix parametrization is defined by $\sigma_{\text{min}} = 4m^2$. It is a natural value in this formulation, since the amplitude is constrained by the elastic unitarity only in the interval $4m^2 \leqslant \sigma_{\q} \leqslant (\sqrt{s}-m)^2$. The advantage of this choice is that it incorporates only the physical intermediate degrees of freedom, and there is no need to regularize the virtual states, as they are absent from the formalism. Moreover, that makes the $B$-matrix framework capable of providing a clear distinction between what one understands by the long-range and short-range effects in the formation of resonances, with the OPE amplitude giving a probability for an exchange of a real, on-shell particle. On the other hand, using a different, lower $\sigma_{\text{min}}$ as in EFT models~\citep{Hansen:2015, Mai:2017b}, pushes the nonphysical singularities associated with this endpoint further away from the physical region.

In particular, the difference in the integration limits between the EFT and $B$-matrix approaches has an important consequence for the behavior of the amplitude below the three-particle threshold when the isobar can form a two-particle bound state, i.e. when $\Fc_{\p}$ has a pole below the two-particle threshold $4m^2$. In this case one would expect to obtain the bound-state--spectator scattering amplitude $\Mc_{22}$ by amputating external interaction amplitudes $\Fc_{\p}$ from $\Mc_{33,\p'\p}$, and setting $\sqrt{\sigma_{\p}}$ and $\sqrt{\sigma_{\p'}}$ equal to the bound-state mass. This can only happen, however, if the integration over the intermediate momentum covers the physical region available to bound-state-spectator states, which is not the case when $\sigma_{\text{min}} \geqslant 4m^2$. When $\sigma_{\text{min}}=4m^2$, the two-particle bound state pole is outside the integration limits, and the resulting amplitude, $\Mc_{22}$ has a wrong two-body threshold behavior.

Let us illustrate this with an example. We consider $S$-wave scattering in the ladder approximation, so that we can drop all angular momentum indices. We define the amputated ladder amplitude 
$\widetilde{\Dc}_{\p'\p}$ by, $\Dc_{\p'\p} = \Fc_{\p'} \widetilde{\Dc}_{\p'\p} \Fc_{\p}$, and it satisfies 
    \beq
    \label{eq:ladder-s-wave}
    & & \widetilde{\Dc}(\sigma_{\p'},s, \sigma_{\p}) = \Gc(\sigma_{\p'}, s, \sigma_{\p}) \nln
    & & + \!\!\!\!\!\int\limits_{\sigma_{\text{min}}}^{\left(\sqrt{s}-m\right)^2} \!\!\!\!\! \frac{d\sigma_{\q}}{2\pi} \, \Gc(\sigma_{\p'}, s, \sigma_{\q}) \, \tau(s,\sigma_{\q})  \, \Fc(\sigma_{\q}) \, \widetilde{\Dc}(\sigma_{\q}, s, \sigma_{\p}) \, .
    \eeq
The $S$-wave projection of the OPE amplitude is given by
    \beq
    \label{eq:OPE-s-wave}
    \Gc(\sigma_{\p'}, s, \sigma_{\p}) = \frac{1}{4 |\p'||\p| } \log \left(\frac{z_{\p' \p} - 1}{z_{\p' \p} + 1}\right) \, ,
    \eeq
with
    \beq
    \label{eq:z-function}
    z_{\p' \p} = \frac{2s\sigma_{\p} - (s+\sigma_{\p} - m^2)(s+m^2 - \sigma_{\p'})}{\lambda^{1/2}(s,\sigma_{\p'}, m^2) \lambda^{1/2}(s,\sigma_{\p},m^2) } 
    \, .
    \eeq
For the two-body, isobar amplitude we use the effective range expansion in the leading order approximation, 
    \beq
    \label{eq:2to2scatt}
    \frac{1}{\Fc(\sigma_{\q})}  = - \frac{1}{a_0} - i \rho(\sigma_{\q}) \, .
    \eeq
The relativistic two-body phase space factor is
    \beq
    \label{eq:phase-space-id}
    \rho(\sigma_{\q}) = \frac{1}{2!} \frac{ q}{8\pi \sqrt{\sigma_{\q}}} = \frac{1}{32 \pi} \sqrt{1 - \frac{4 m^2}{\sigma_{\q}}}  \, ,
    \eeq
while the dimensionless parameter $a_0$ is related to the scattering length. Near the bound-state pole, which is determined by condition $\Fc(\sigma_b)^{-1} = 0$, the isobar amplitude is
    \beq
    \label{eq:2pole}
    \Fc(\sigma_{\q}) = - \frac{g^2}{\sigma_{\q} - \sigma_b + i\epsilon} \, ,
    \eeq
with the pole position and the residue given by
    \beq
    \label{eq:pole-energy}
    \sigma_b = \frac{4m^2}{1+\left( \frac{32\pi}{a_0} \right)^2} \, , ~~~
    g = \frac{32 \pi}{ \sqrt{2 a_0} } \frac{\sigma_b}{m} \, ,
    \eeq
respectively. The contribution from the imaginary part $\im  \Fc(\sigma_{\q}) = - \pi \, \delta(\sigma_{\q} - \sigma_b)$ to the integral in Eq.~\eqref{eq:ladder-s-wave} in the limit $\sigma_{\p'}, \sigma_{\p} \to \sigma_b$, reproduces the bound-state--spectator phase space factor
    \beq
    \label{eq:phase-space-diff}
    \rho_2(s) = \frac{\lambda^{1/2}(s,\sigma_b,m^2)}{16 \pi s} \, .
    \eeq
However, this is not the case if $\sigma_{\text{min}} = 4m^2 > \sigma_b$, for which the two-particle unitarity constraint on the imaginary part is not reproduced. A similar argument can be made in the low-energy approximation, see App.~\ref{app:B}, where a comparison of the $B$-matrix formalism with the non-relativistic EFT is presented.

The $B$-matrix is constructed to respect only the direct-channel unitarity while being nescient about left-hand singularities. Thus to describe bound-state--spectator scattering, such a state, or any other particle below the three-particle threshold, has to be included explicitly. For this reason, the $B$-matrix description of $\3\to\3$ scattering alone cannot be directly compared, for example, with the recent calculations of the bound-state--spectator scattering length in Refs.~\citep{Romero-Lopez:2019, Jackura:2020bsk}. In the following sections we show how resolve these issues and first give the multi-channel generalization.

\hspace{3pt}

\section{Multi-channel formalism}
\label{sec:multi-channel}

We consider the  $\2\to\2$, $\3\to\2$, $\2\to\3$ and $\3\to\3$ scattering processes. The two-particle state contains a bound state of mass $M<2m$, formed by the isobar appearing in the three-particle channel, and the spectator of mass $m$. For simplicity, the particles are taken to be scalars and distinguishable. We introduce the $\n\to\m$ amplitudes $\Mc_{mn}$, which are matrix elements of the $T$ matrix describing different reaction channels. Their precise definition is provided in App.~\ref{app:C}. As in the elastic $\3\to\3$ case, the amplitude $\Mc_{33}$ has both a connected and disconnected parts, while $\Mc_{32}$, $\Mc_{23}$, $\Mc_{22}$ are connected by definition. Therefore one can consistently write $\Mc_{nm} = \Ac_{nm}$ for both $n,m\neq3$. The three-particle states are described using the same kinematic variables $(\p\ell m_\ell)$ described in App.~\ref{app:A}. In this basis partial wave projected $\2\to\3$ amplitude is a ``vector" (meaning it depends only on angular momentum of one external isobar) $\Ac_{32,\p'}$, while the $\2\to\2$ amplitude $\Ac_{22}$ is a ``scalar". The two-body system of the bound-state and spectator is described by the total invariant mass squared $s$ or the relative momentum $\k$ between the particles in the center of mass frame given by 
    \beq
    |\k| =  \frac{1}{2\sqrt{s}} \lambda(s,M^2,m^2) .
    \eeq
The angular orientation $\Omega_{\hat{\k}}$ of the outgoing spectator's momentum in the two-body system is defined with respect to the incoming spectator's momentum, either in the two-body or three-body state.

\begin{figure}[b!]
    \centering
    \includegraphics[ width=0.45\textwidth]{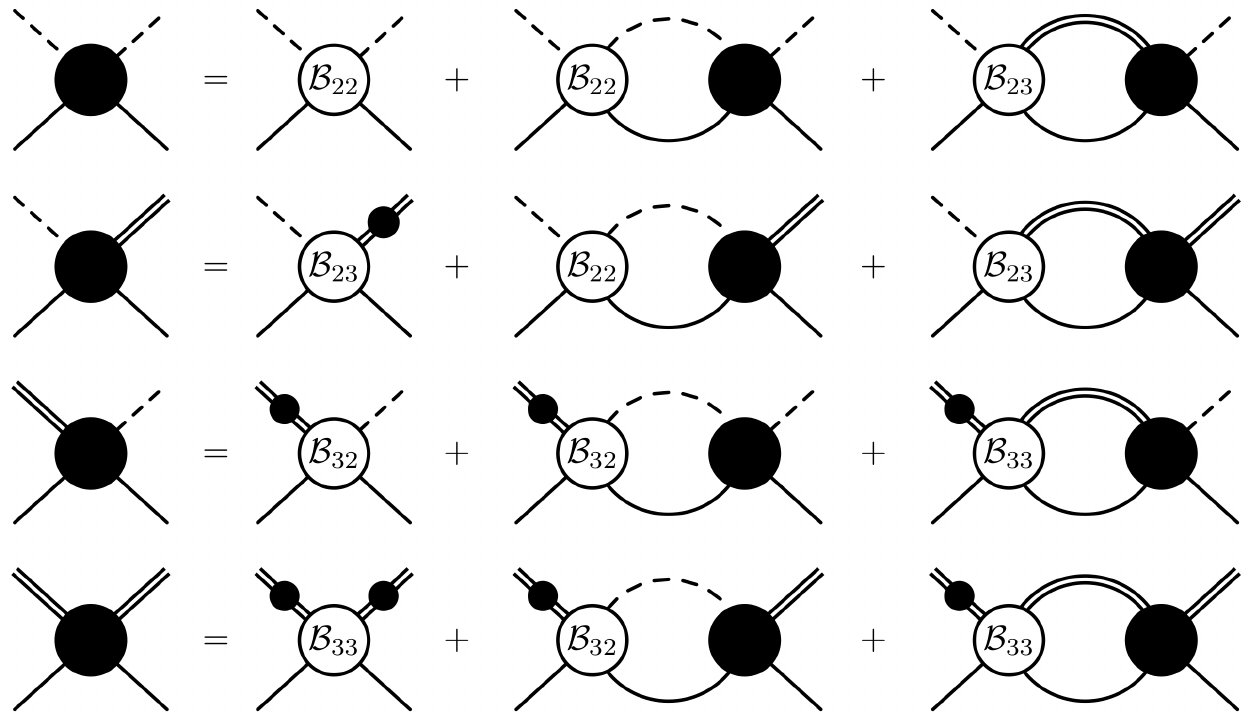}
    \caption{Diagrammatic representation of the multi-channel $B$-matrix framweork, Eqs.~\eqref{eq:b-matrix-22}--\eqref{eq:b-matrix-33}. Amplitudes $\Ac_{mn}$ are represented by solid circles and can be differentiated from the different types of external legs. Dashed line represents the two-body bound state of mass $M$. The $\Bc_{33}$ kernel is decomposed as in Fig.~\ref{fig:b-matrix}, while other, real kernels describe just short-range interactions.}
    \label{fig:b-matrix-multi-channel}
\end{figure}

Each amplitude has its corresponding $B$-matrix kernel. These are real functions $\Bc_{22} \equiv \Bc_{22}(s,\hat{\k})$, $\Bc_{23,\p} \equiv \Bc_{23, \ell m_\ell}(s,\sigma_{\p},\hat{\k})$ and $\Bc_{32,\p'} \equiv \Bc_{32,\ell' m_\ell'}(\sigma_{\p'},s,\hat{\k})$, unconstrained by unitarity. The $\Bc_{33,\p'\p}$ kernel was defined in the previous section and in App.~\ref{app:A}. Denoting integration over the implicit angular dependence by $ \int_{\hat{\k}} = \int \frac{d\Omega_{\hat{\k}}}{4\pi}$, the generalized $B$-matrix 
parameterization of the connected amplitudes $\Ac_{mn}$ is given by
\begin{widetext}
    \beq
    \label{eq:b-matrix-22}
    \Ac_{22} &=& \Bc_{22} + \int_{\hat{\k}} \Bc_{22} i \rho_2 \Ac_{22} + \int_{\q} \Bc_{23,\q} \,  \Ac_{32,\q} , \\
    \label{eq:b-matrix-23}
    \Ac_{23,\p} &=& \Bc_{23, \p} \, \Fc_{\p} + \int_{\hat{\k}} \Bc_{22} i \rho_{2} \Ac_{23,\p} + \int_{\q} \Bc_{23,\q} \, \Ac_{33,\q \p } , \\
    \label{eq:b-matrix-32}
    \Ac_{32,\p'} &=& \Fc_{\p'} \, \Bc_{32,\p'} +  \int_{\hat{\k}} \Fc_{\p'} \Bc_{32,\p'} i \rho_2 \Ac_{22} + \int_{\q} \Fc_{\p'} \, \Bc_{33,\p'\q} \, \Ac_{32,\q} , \\
    \label{eq:b-matrix-33}
    \Ac_{33,\p'\p} &=& \Fc_{\p'} \, \Bc_{33,\p'\p} \, \Fc_{\p} + \int_{\hat{\k}} \Fc_{\p'} \, \Bc_{32,\p'} i \rho_{2} \Ac_{23,\p} + \int_{\q} \Fc_{\p'} \, \Bc_{33,\p'\q} \, \Ac_{33,\q\p} \ .
    \eeq
\end{widetext}
As shown in App.~\ref{app:C}, the amplitudes given above satisfy unitarity above the three-body threshold $s_{\text{th},3} = (3m)^2$. It is important to note that this representation in general does not satisfy the unitary between the bound-state--particle threshold, $s_{\text{th},2} = (M+m)^2$ and the three-body threshold $s_{\text{th},3}$. This is because the three-body channel can contribute a nonzero imaginary part of the amplitude $\Ac_{mn}$ below $s_{\text{th},3}$. The diagrammatic representation of the above equations is shown in Fig.~\ref{fig:b-matrix-multi-channel} and the formalism can be easily generalized to include other channel.

Equations~\eqref{eq:b-matrix-22}--\eqref{eq:b-matrix-33} involve mixing between different channels. As can be seen, amplitudes $\Ac_{22}$ and $\Ac_{32,\p'}$ depend on each other, but not on $\Ac_{23}$ or $\Ac_{33}$. However, it would be incorrect to infer that the $B$-matrix representation presented above does not couple the $\3 \to \3$ and $\2\to\2$ physics, and is two pairs of independent linear equations. The physical content of the dynamics is contained in $\Bc_{nm}$ kernels, which are shared by the seemingly uncoupled equations. This can be seen clearly when formally decoupling them and solving for the individual amplitudes. Firstly, in an analogy with the $\3\to\3$ formalism, the isobar-amplitude is amputated in the three-body channels containing isobars, defining $\Ac_{32,\p'} = \Fc_{\p'} \, \widetilde{\Ac}_{32,\p'}$ etc. Although there is no amputation needed for the bound-state--particle channels, $\Ac_{22} = \widetilde{\Ac}_{22}$ is written to maintain consistency in the notation. Having the equivalents to Eqs.~\eqref{eq:b-matrix-22}--\eqref{eq:b-matrix-33} for the amputated amplitudes, one can eliminate $\widetilde{\Ac}_{23,\p}$, by transforming Eq.~\eqref{eq:b-matrix-23} to a form
    \beq
    \label{eq:rewritten-23}
    \widetilde{\Ac}_{23,\p} = \frac{1}{1 - i \rho_2 \Bc_{22}} \left[ \Bc_{23,\p} + \int_{\q} \Bc_{23,\q} \, \Fc_{\q} \, \widetilde{\Ac}_{33,\q\p} \right]  .~~
    \eeq
And using the above result in Eq.~\eqref{eq:b-matrix-33} one obtains
    \beq
    \label{eq:decoupled-33}
    \widetilde{\Ac}_{33,\p'\p}
    &=& \Hc_{33,\p'\p} +  \int_{\q} \Hc_{33,\p'\q} \, \Fc_{\q} \, \widetilde{\Ac}_{33,\q\p} \, ,
    \eeq
where the effective three-body kernel is 
    \beq
    \label{eq:modified-kernel-33}
    \Hc_{33,\p'\p} \equiv \Bc_{33,\p'\p} + \frac{\Bc_{32,\p'} i \rho_2 \Bc_{23,\p} }{1 - i \rho_2 \Bc_{22}} \, ,
    \eeq
and it includes the all-orders coupling of the three particle state to the two-particle state containing the bound-state and the spectator. This interaction is not present in the elastic $\3\to\3$ $B$-matrix formalism and thus has to be included explicitly via Eq.~\eqref{eq:rewritten-23} above. Analogously, using Eq.~\eqref{eq:b-matrix-22} in Eq.~\eqref{eq:b-matrix-32} one obtains the integral equation for the $\widetilde{\Ac}_{32}$ amplitude, 
    \beq
    \label{eq:decoupled-32}
    \widetilde{\Ac}_{32,\p'} = \Hc_{32,\p'} + \int_{\q} \Hc_{33,\p'\q} \, \Fc_{\q} \, \widetilde{\Ac}_{32,\q} \ ,
    \eeq
where
    \beq
    \label{eq:modified-kernel-32}
    \Hc_{32,\p'} = \Bc_{32,\p'} + \frac{\Bc_{32,\p'} i \rho_2 \Bc_{22}}{1-i \rho_2 \Bc_{22}} \ .
    \eeq
To obtain decoupled equations for the $\widetilde{\Ac}_{22}$ and $\widetilde{\Ac}_{23}$ one would have to solve Eqs.~\eqref{eq:decoupled-33} and \eqref{eq:decoupled-32} first. Assuming that they are generalized matrix equations both in the angular momentum space and the continuous momentum space, one can write down their formal solutions as
    \beq
    \label{eq:formal-solution-33}
    \widetilde{\Ac}_{33}
    &=& [\one -  \Hc_{33} \, \Fc]^{-1} \Hc_{33} , \\
    \label{eq:formal-solution-32}
    \widetilde{\Ac}_{32} &=& [\one -  \Hc_{33} \, \Fc]^{-1}\Hc_{32} .
    \eeq
Then the remaining amplitudes are
    \beq
    \label{eq:formal-solution-22}
    \widetilde{\Ac}_{22} &=& \frac{1}{1- i \rho_2 \Bc_{22}} \! \left[ \Bc_{22} + \Bc_{23} \, \Fc \, [\one -  \Hc_{33} \, \Fc]^{-1}  \Hc_{32}  \right] \! , ~~~\\
    \label{eq:formal-solution-23}
    \widetilde{\Ac}_{23} &=& \frac{1}{1- i \rho_2 \Bc_{22}} \! \left[ \Bc_{23} + \Bc_{23} \, \Fc [\one -  \Hc_{33} \, \Fc]^{-1}  \Hc_{33}  \right] \! . ~~~
    \eeq
As can be seen $\Bc_{33}$ enters the solution for $\wAc_{22}$ through $\Hc_{33}$ in the denominator term in Eq.~\eqref{eq:formal-solution-22} and affects the two-body physics as long as $\Bc_{23}$ and $\Bc_{32}$ are nonzero. The result can be also expressed in a more concise form by directly solving the generalized matrix equivalents of Eqs.~\eqref{eq:b-matrix-22} and \eqref{eq:b-matrix-32}, which leads to, 
    \beq
    \label{eq:formal-solution-22-2}
    \wAc_{22} &=& [\one -\Hc_{22} i \rho_2]^{-1} \Hc_{22} \, , \\
    \label{eq:formal-solution-23-2}
    \wAc_{23} &=& [\one -\Hc_{22} i \rho_2]^{-1} \Hc_{23} \, . 
    \eeq
where
    \beq
    \label{eq:modified-kernel-22}
    \Hc_{22} = \Bc_{22} + \Bc_{23} \Fc [\one - \Bc_{33} \Fc ]^{-1} \Bc_{32} \, , \\
    \label{eq:modified-kernel-23}
    \Hc_{23} = \Bc_{23} + \Bc_{23} \Fc [\one - \Bc_{33} \Fc ]^{-1} \Bc_{33} \, .
    \eeq
The $\2\to\2$ scattering can occur even in the absence of the direct interactions between the bound-state and the spectator, i.e. for $\Bc_{22}=0$. In this case, the dynamics of the two-body scattering is described entirely by the physics involving the three particles. The solution for the $\wAc_{33}$ amplitude is governed by the $\Hc_{33}$ kernel, which contains the direct interaction kernel, $\Bc_{33}$, and an effective interaction due to mixing with the bound-state-spectator intermediate state. Potential problems with the analytic continuation of the single-channel $\3 \to \3$ scattering with two-body resonances were already discussed in Ref.~\citep{Jackura:2018}. In the following section, we discuss the analytical properties in the case under study, i.e. in the presence of two-body bound states.

\section{Short-range interactions Model}
\label{sec:contact}

We turn to investigation of the generalized $B$-matrix parametrization presented in Eqs.~\eqref{eq:b-matrix-22}–\eqref{eq:b-matrix-33} within the contact-interaction model, i.e. with the effects of long-range interactions being neglected. The OPE amplitude is set to zero, $\Gc=0$, in which case $\Bc_{33,\p'\p} = \Rc_{\p'\p}$, and the short-range kernels $\Rc_{33,\p'\p}$ and $\Bc_{23,\p}$, $\Bc_{32,\p'}$, $\Bc_{22}$ are assumed to be momentum independent. They are rewritten as a set of real coupling constants $\Rc_{33} = g_{33}$,  $\Bc_{23} = g_{23} = \Bc_{32} = g_{32}$, $\Bc_{22} = g_{22}$. We consider only the $S$-wave, and amputate the external isobars interactions, which allows to obtain equations for the amplitudes,
    \beq
    \label{eq:contact-equation22}
    \widetilde{a}_{22}(s) &=& g_{22} 
    + g_{22} i \rho_2(s) \, \widetilde{a}_{22}(s) \nln
    & & + g_{32} \int_{\q} \Fc(\sigma_{\q}) \, \widetilde{a}_{32}(\sigma_{\q}, s) \, , \\
    \label{eq:contact-equation32}
    \widetilde{a}_{32}(\sigma_{\p'},s) &=& g_{32} 
    +  g_{32} i \rho_2(s) \, \widetilde{a}_{22}(s) \nln
    & & + g_{33} \int_{\q} \Fc(\sigma_{\q}) \, \widetilde{a}_{32}(\sigma_{\q}, s) \, , \\
    \label{eq:contact-equation23}
    \widetilde{a}_{23}(s, \sigma_{\p}) &=& g_{32} 
    +  g_{22} i \rho_2(s) \, \widetilde{a}_{23}(s,\sigma_{\p}) \nln
    & & + g_{32} \int_{\q} \Fc(\sigma_{\q}) \, \widetilde{a}_{33}(\sigma_{\q}, s, \sigma_{\p}) \, , \\
    \label{eq:contact-equation33}
    \widetilde{a}_{33}(\sigma_{\p'},s, \sigma_{\p}) &=& g_{33} 
    +  g_{32} i \rho_2(s) \, \widetilde{a}_{23}(s,\sigma_{\p}) \nln
    & & + g_{33} \int_{\q} \Fc(\sigma_{\q}) \, \widetilde{a}_{33}(\sigma_{\q}, s, \sigma_{\p}) \, ,
    \eeq
where $\widetilde{a}$ was used to differentiate the amplitudes in this approximation from the more general case discussed in the previous section. One can immediately find uncoupled equations for $\widetilde{a}_{33}$ and $\widetilde{a}_{33}$ using Eqs.~\eqref{eq:decoupled-33} and \eqref{eq:decoupled-32},
    \beq
    \label{eq:contact-sol33}
    \widetilde{a}_{33}(\sigma_{\p'},s, \sigma_{\p}) &=& h_{33}(s) \nln
    & & +  h_{33}(s) \int_{\q} \Fc(\sigma_{\q}) \, \widetilde{a}_{33}(\sigma_{\q}, s, \sigma_{\p}) \, ,
    \eeq
and
    \beq
    \label{eq:contact-sol32}
    \widetilde{a}_{32}(\sigma_{\p'},s) &=& h_{32}(s) + h_{33}(s) \! \int_{\q} \Fc(\sigma_{\q}) \, \widetilde{a}_{32}(\sigma_{\q}, s) \, ,~~ 
    \eeq
where $h_{33}$ and $h_{32}$ follow from Eqs.~\eqref{eq:modified-kernel-33} and \eqref{eq:modified-kernel-32},
    \beq
    h_{33}(s) &=& g_{33} + \frac{g_{32}^2 i \rho_2(s) }{1 - g_{22} i \rho_2(s) } \, , \\
    h_{32}(s) &=& g_{32} + \frac{ g_{22} g_{32} i \rho_2(s)}{1-g_{22} i \rho_2(s)} \, .
    \eeq
Both equations can be solved by noticing that their RHS do not depend on the left argument $\sigma_{\p'}$, meaning the amplitudes on the LHS, $\widetilde{a}_{33}(\sigma_{\p'}, s, \sigma_{\p}) = \widetilde{a}_{33}( s, \sigma_{\p})$ and $\widetilde{a}_{32}(\sigma_{\p'}, s) = \widetilde{a}_{32}(s)$. Furthermore, in Eq.~\eqref{eq:contact-sol33}, the $s$ and $\sigma_{\p}$ dependence can be factorized, eventually eliminating $\sigma_{\p}$ dependence from $\widetilde{a}_{33}$. In consequence, both amplitudes can be moved outside of the corresponding integrals, yielding algebraic equations with solutions,
    \beq
    \label{eq:d33sol}
    \widetilde{a}_{33}(s) &=& \frac{h_{33}(s)}{1 - h_{33}(s) \Ic(s) } \nln
    &=& \frac{g_{33} + G i \rho_2(s) }{1 - g_{22} i \rho_2(s) - [g_{33} + G \, i \rho_2(s) ] \, \Ic(s) } \, , \\
    \label{eq:d32sol}
    \widetilde{a}_{32}(s) &=& \frac{h_{32}(s)}{1 - h_{33}(s) \Ic(s) } 
    \nln
    &=& \frac{g_{32}}{1 - g_{22} i \rho_2(s) - [g_{33} + G \, i \rho_2(s) ] \, \Ic(s) } \, ,
    \eeq
where $G \equiv g_{32}^2 - g_{33} g_{22}$, and  
    \beq
    \label{eq:integral}
    \Ic(s) &=& \int\limits_{\sigma_{\text{min}}}^{(\sqrt{s}-m)^2} \frac{d\sigma_{\q}}{2\pi} \, \tau(s,\sigma_{\q}) \, \Fc(\sigma_{\q}) \nln
    &=& \frac{1}{16\pi^2 s} \!\!\! \int\limits_{\sigma_{\text{min}}}^{(\sqrt{s}-m)^2} \!\!\! d\sigma_{\q} \ \Jc(\sigma_{\q},s)  .
    \eeq
As can be seen, the above solutions are special cases of the formal solutions in Eqs.~\eqref{eq:formal-solution-33} and \eqref{eq:formal-solution-32}. Analogous arguments lead to solutions for the remaining amplitudes,
    \beq
    \label{eq:d23sol}
    \widetilde{a}_{23}(s) &=& \frac{g_{32} }{1 - g_{22} i \rho_2(s) - [g_{33} + G \, i \rho_2(s) ] \, \Ic(s) } \, , \\
    \label{eq:d22sol}
    \widetilde{a}_{22}(s) &=& \frac{g_{22} + G \, \Ic(s)}{1 - g_{22} i \rho_2(s) - [g_{33} + G \, i \rho_2(s) ] \, \Ic(s) } \, .
    \eeq
To make our considerations more concrete, from now on we take the model of Eq.~\eqref{eq:2to2scatt} for the isobar interaction amplitude $\Fc(\sigma_{\q})$, which results in the integrand in Eq.~(\ref{eq:integral}) given by
    \beq
    \label{eq:integrand}
    \Jc(\sigma, s) = \frac{\sqrt{\sigma - [\sqrt{s}-m]^2}\sqrt{\sigma - [\sqrt{s}+m]^2}}{-\frac{1}{a_0} + \frac{1}{32\pi} \sqrt{\frac{4m^2}{\sigma} - 1 }} \, .
    \eeq
    
In addition to the canonical $B$-matrix model, in which $\sigma_{\text{min}} = 4m^2$ and couplings can take arbitrary values, we also distinguish the ``EFT-like" contact interaction model, which imitates the three-body EFT approaches by including the virtual momenta in the integration. It is introduced in order to emphasize the consequences of the different choices of $\sigma_{\text{min}}$. We denote the EFT-like kernel, which is defined by the setting $\sigma_{\text{min}}=0$ in Eq.~\eqref{eq:integral}, as $\Ic_{\text{EFT}}$. It can be shown that below the three-body threshold $s_{\text{th},3}$,
    \beq
    \label{eq:EFTkernel}
    \im \Ic_{\text{EFT}}(s) = g^2 \, \rho_2(s) \, \theta(s-s_{\text{th},2}) \, ,
    \eeq
i.e. $\im \, \Ic_{\text{EFT}}$ behaves as the two-body phase space multiplied by the isobar residue of Eq.~\eqref{eq:pole-energy}. In this model couplings $g_{22}=g_{23}=0$ and the three-body coupling is renamed as $g_{33,\text{EFT}}$. Thus, below the three-body threshold the EFT-like amplitude is given by
    \beq
    \label{eq:d33solEFT2}
    \widetilde{a}_{33,\text{EFT}} &=& \frac{1}{g^2} \frac{1}{\frac{1}{g^2 g_{33,\text{EFT}}  } - i \rho_2(s) } \ .
    \eeq
The corresponding $\2\to\2$ and $\2\to\3$ amplitudes are obtained by removing the residues of the external isobar amplitudes, viz., $\widetilde{a}_{22,\text{EFT}} = g^2 \, \widetilde{a}_{33,\text{EFT}}$, and $\widetilde{a}_{32,\text{EFT}} = g \, \widetilde{a}_{33,\text{EFT}}$.

Unitarity relations summarized in Eqs.~\eqref{eq:unitarity-22}--\eqref{eq:unitarity-33} imply,
    \beq
    \label{eq:unitarity-check}
    \im \widetilde{a}_{22}(s) &=& \rho_2(s) \, |\widetilde{a}_{22}(s)|^2 \, \theta(s-s_{\text{th},2}) \nln
    & & + \im \, \Ic(s) \, |\widetilde{a}_{32}(s)|^2 \, \theta(s - s_{\text{th},3}) \, , \\
    \label{eq:unitarity-check32}
    \im \widetilde{a}_{32}(s) &=& \rho_2(s) \, \widetilde{a}_{32}^*(s) \widetilde{a}_{22}(s) \, \theta(s-s_{\text{th},2}) 
    \nln
    & & + \im \Ic(s) \, |\widetilde{a}_{32}(s)|^2 \, \theta(s - s_{\text{th},3}) \, , \\
    \label{eq:unitarity-check33}
    \im \widetilde{a}_{33}(s) &=& \rho_2(s) \, |\widetilde{a}_{32}(s)|^2 \, \theta(s-s_{\text{th},2}) \nln
    & & + \im \Ic(s) \, |\widetilde{a}_{33}(s)|^2 \, \theta(s - s_{\text{th},3}) \, .
    \eeq
with $\2\to\3$ case being identical to the $\3\to\2$ one. Because of the step function $\theta(s-s_{
\text{th},3})$ the contribution to the RHS from the three-body channel ought to vanish below $s_{\text{th},3}$. However, by directly calculating the imaginary part of Eq.~\eqref{eq:d22sol} one obtains,
    \beq
    \label{eq:unitarity-check}
    \im \widetilde{a}_{22}(s) &=& \rho_2(s) \, |\widetilde{a}_{22}(s)|^2 + \im \, \Ic(s) \, |\widetilde{a}_{32}(s)|^2 \, .
    \eeq
This agrees with above equations for $s \geqslant s_{\text{th},3}$, but disagrees, if $\im \, \Ic(s)$ is nonzero for $s_{\text{th},3} > s \geqslant s_{\text{th},2}$.

\begin{figure}[ht!]
\begin{center}
\subfigure[~EFT-like model with cutoff $\sigma_\min/m^2 = 0$. ]
{
\includegraphics[width=0.42\textwidth]{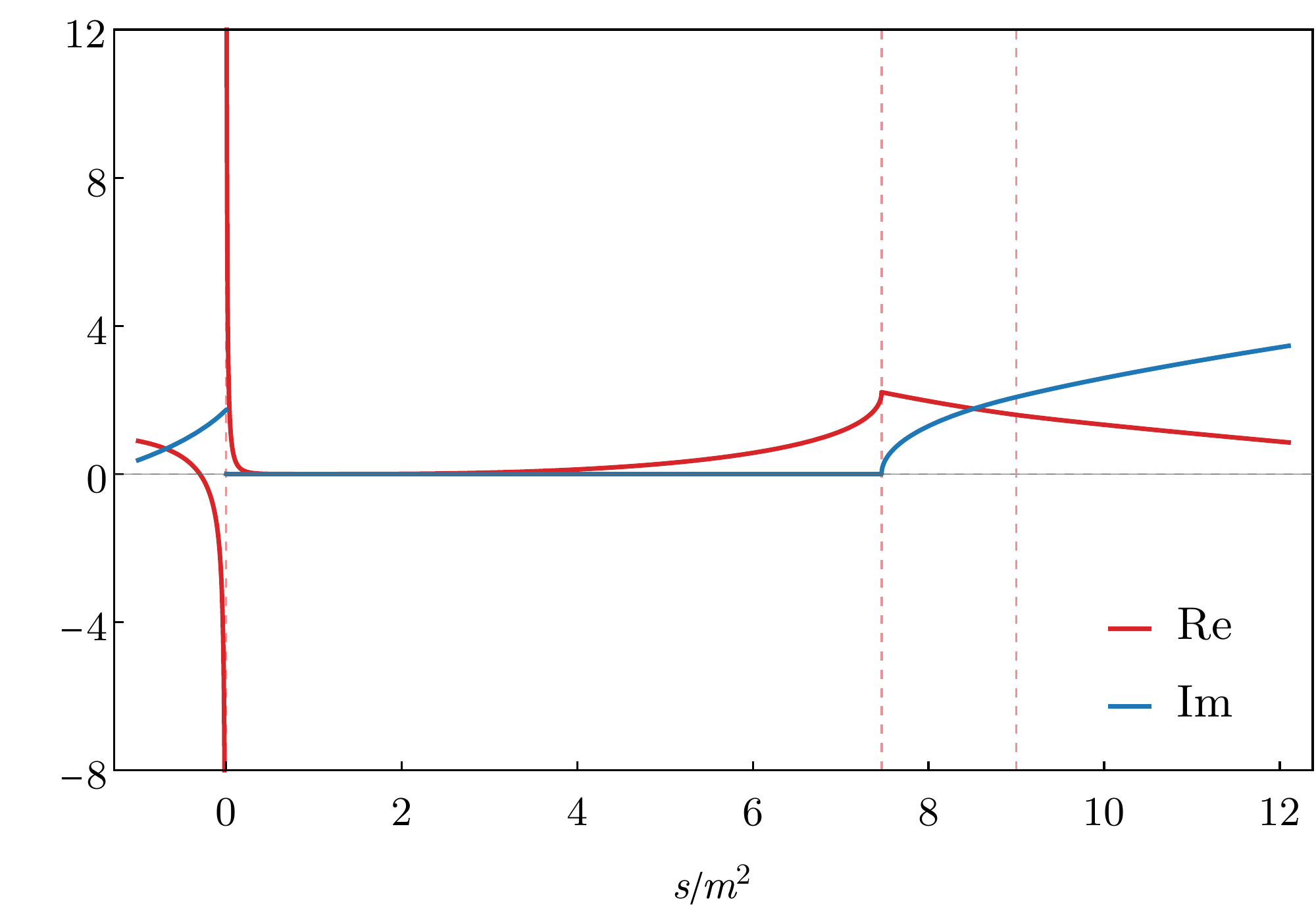}
\label{fig:functIa}
}
\subfigure[~B-matrix model with the physical limit of integration $\sigma_\min/m^2 = 4$. ]
{
\includegraphics[width=0.42\textwidth]{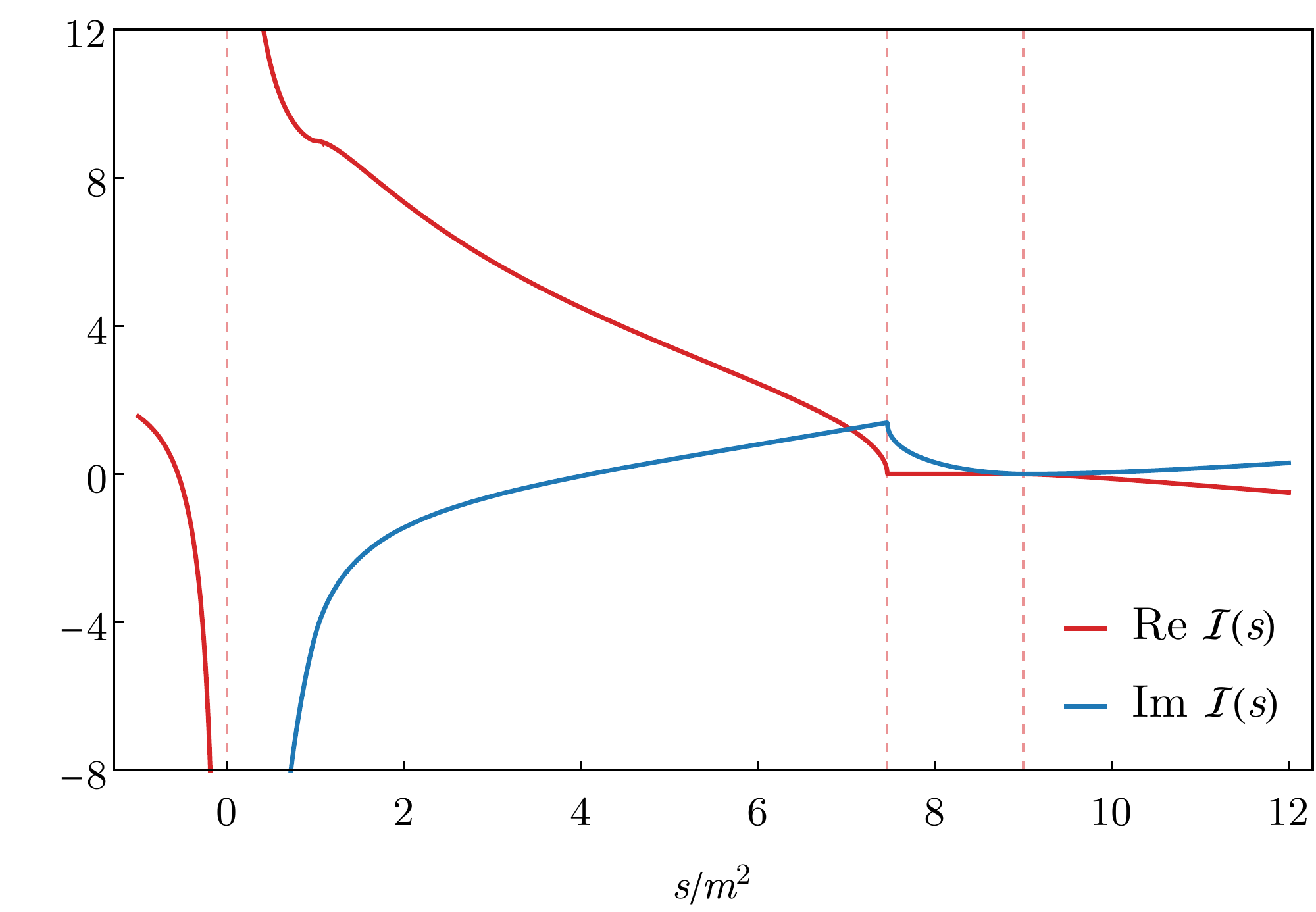}
\label{fig:functIb}
}
\end{center}
\caption{The kernel $\Ic(s)$ for two different choices of $\sigma_\min$. The bound state has a mass $M^2 = 3m^2$ which corresponds to the bound-state--spectator threshold energy $s_{\text{th},2} \approx 7.456 m^2$. For $s=0$ the singularity from the definition of the three-body phase space factor $\tau(\sigma,s)$ occurs. The three-body threshold occurs at $s_{\text{th},3}=(3m)^2$. The points of non-analyticity are highlighted by the dashed, red lines.}
\label{fig:functI}
\end{figure}

We can now investigate the analytic properties of the amplitudes. The integrand $\Jc(\sigma,s)$ considered as a function of $\sigma$, i.e. for fixed values of $s$, has a singularity structure characterized by five points: a left-hand cut branch point at $\sigma_0 = 0$, a fixed two-body bound state pole at $\sigma_b = M^2$, a two-body threshold branch point at $\sigma_2 = (2m)^2$, and two $s$-dependent branch points $\sigma_3(s) = (\sqrt{s} - m)^2$ and $\sigma_4(s) = (\sqrt{s} + m)^2$. 
Since $\sigma_{3,4}$ depend on $s$, the position and number of singular points changes with $s$. Namely, at the three-particle threshold $s = s_{\text{th},3}$ points $\sigma_2$ and $\sigma_3$ coincide. For $s = s_{\text{th},2}$ the singularity at $\sigma_3$ coincides with the location of the bound-state pole $\sigma_3=\sigma_b$, while for $s = m^2$, $\sigma_0 = \sigma_3$ and $\sigma_2 = \sigma_4$. Finally, for $s=0$ points $\sigma_{3}$ and $\sigma_{4}$ coincide, and as $s$ becomes negative and decreases, they become complex and move away from the real axis. The change in behavior of $\Jc$ for decreasing values of $s$ is illustrated in Fig.~\ref{fig:functJflow} in App.~\ref{app:D}. The trajectories of the singularities $\sigma_3$ and $\sigma_4$ as functions of $s$ are shown in Fig.~\ref{fig:mov_sin}.
 
In general, singularities of $\Ic$ can appear for those values of $s$ for which: a) one of the singularities in $\sigma$ discussed above, coincides with the lower limit of integration, $\sigma_\min$; b) one of the singularities in $\sigma$ discussed above, coincides with the upper limit of integration, $(\sqrt{s}-m)^2$; c) two movable singularities of $\Jc(\sigma,s)$, pinch the integration contour, which in our case is the real line interval between $\sigma_\min$ and $(\sqrt{s}-m)^2$. The plot of $\Ic$ for two choices of the lower limit, i.e. for $\sigma_{\text{min}}=4m^2$, which is canonical for the $B$-matrix formalism and $\sigma_{\text{min}}=0$, which gives the EFT-like model, is shown in Fig.~\ref{fig:functI}.

We shall discuss these two cases separately, starting with the EFT-like model. At $s=0$ there is a pole originating from the three-body phase space. Then there is the right-hand branch cut that starts at $s=s_{\text{th},2}$. It originates form the two-body bound state pole at $\sigma=M^2$ colliding with the upper limit of integration (condition b). The three-body branch cut starts at $s_{\text{th,3}}$, and originates from the upper integration limit coinciding with $\sigma_2=(2m)^2$ (condition b). In total we find one pole and two branch cuts. The cuts are physical (to the right) and are associated with two and three body thresholds.

Now we consider the case when $\sigma_{\text{min}} = 4 m^2$. There is the same pole at $s=0$ and the right hand cut associated with the three-body threshold as in the EFT-like model. In addition, however,  nonphysical branch points appear. Instead of the right-hand cut associated with the two-body threshold, $s_{\text{th},2}$ there is now a left-cut starting at this point. The direction of the cut is different than in the EFT-like model because the direction of the integration in Eq.~\eqref{eq:integral} changes, since $(\sqrt{s_{\text{th},2}}-m)^2 < \sigma_{\text{min}}$. Additionally a left-hand cut appears at $s_{\text{th},3}$ from the lower integration limit $\sigma_{\text{min}}$ colliding with the movable singularity $\sigma_3$ (condition a).
  
\begin{figure}[t!]
\begin{center}
\subfigure[~Undispersed model amplitude $\widetilde{a}_{33}(s)$]
{
\includegraphics[width=0.42\textwidth]{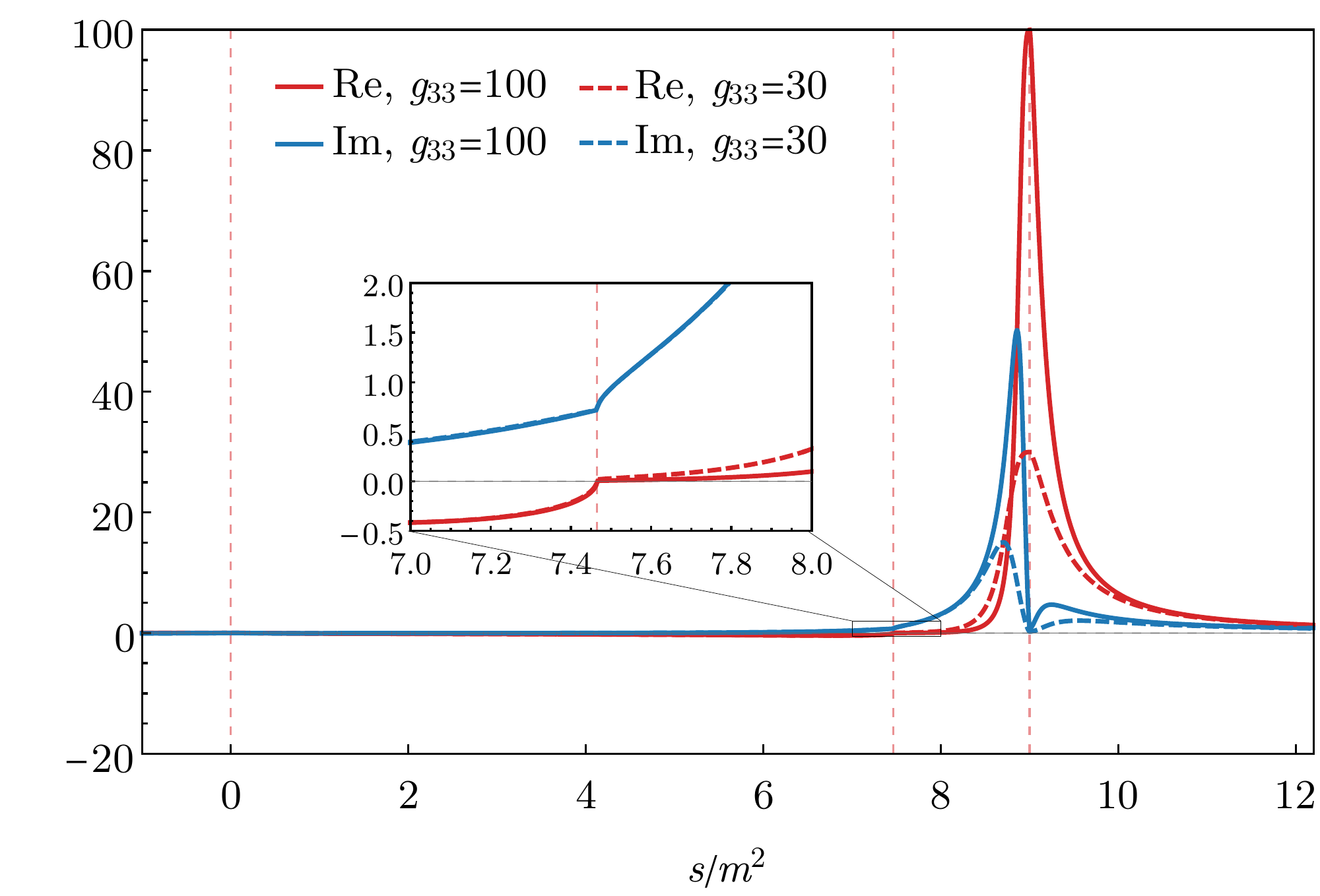}
\label{fig:amp_a}
}
\subfigure[~Dispersed model amplitude $\widetilde{a}_{33,d}(s)$ ]
{
\includegraphics[width=0.42\textwidth]{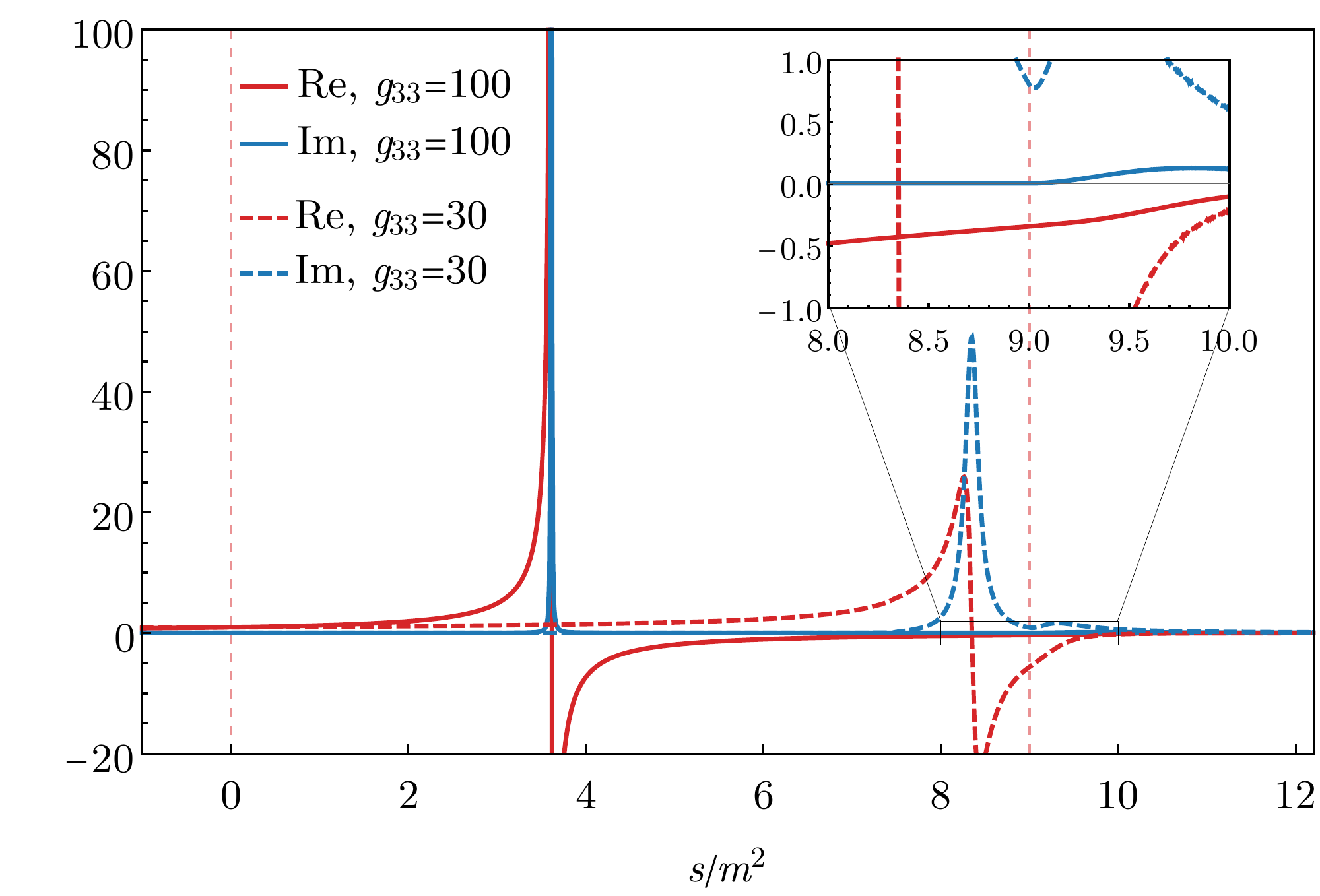}
\label{fig:amp_b}
}
\end{center}
\caption{The comparison between the $\3\to\3$ amplitudes in the model with undispersed (a) and dispersed (b) integral kernel $\Ic(s)$. The value of the couplings are $g_{22}=1$, $G=0$ and $g_{33}=100$ or $g_{33}=30$. The two-body bound-state pole is at $M^2=3m^2$, which corresponds to the bound-state--spectator threshold energy $s_{\text{th},2} \approx 7.456 m^2$. The three-body threshold is at $s=9m^2$. The points of non-analyticity are highlighted by the dashed, red lines. As can be seen, the original model is insensitive to the changes of $g_{33}$ and has a nonphysical branch point at $s_{\text{th},2}$. The dispersed model $\widetilde{a}_{33,d}(s)$ has no singularities below the three-body threshold other than the three-body bound-state pole at $s_{*}/m^2 \approx 3.618$ for large enough $g_{33}$. For coupling which is too small the pole does not appear on the real axis.}
\label{fig:amp100}
\end{figure}

The nonphysical left-hand cuts of $\Ic(s)$ have dire consequences for the scattering amplitude. In the $B$-matrix model, with $\sigma_{\text{min}} = 4m^2$, the imaginary part of $\Ic$ is nonzero between the two thresholds $s_{\text{th},2}$ and $s_{\text{th},3}$, and as a consequence the $\widetilde{a}_{22}$ amplitude does not satisfy the two-body unitarity relation of Eq.~\eqref{eq:unitarity-check}, i.e. $\im \, [\widetilde{a}_{22}(s)]^{-1} \ne - \rho(s)$. Furthermore, when the three-body interactions are decoupled from the bound-state--particle channel, $g_{32}=g_{22}=0$, one finds
    \beq
    \label{eq:simple-d33}
    \widetilde{a}_{33}(s) = \frac{1 }{\frac{1}{g_{33}} - \Ic(s)  }. 
    \eeq
The location of a three-body bound state is given by conditions  
    \beq
    \label{eq:simple-d33-pole}
    \re \Ic(s_p) = \frac{1}{g_{33}} \ , ~~~ \im \Ic(s_p) = 0 \ ,
    \eeq
and can happen of a single value of $g_{33}$, since as shown in Fig.~\ref{fig:functIb}, $\im\, \Ic(s)$ vanishes only for one value of $s$ below the the $s_{\text{th},3}$ threshold. This is in contrast to what happens in the EFT-like model, where, as can be seen in Fig.~\ref{fig:functIa}, there is a finite interval of energies for which $\im \, \Ic_{\text{EFT}}(s)$ vanishes. In other words, in the $B$-matrix model, once the two-body scattering length $a_0$ is fixed, the control over the only parameter of the three-body physics, such as $g_{33}$, becomes illusory, as it does not affect the physical predictions of the model. This can be seen in the Fig.~\ref{fig:amp_a}, where the amplitude $\widetilde{a}_{33}$ shows only a nonphysical bump, which scales with $g_{33}$, and a spurious singularity occurring at $s_{\text{th},2}$.

The $B$-matrix formalism is the extension of the $K$-matrix approach to a three-body channel. What was found here is that this simple extension results in numerous spurious singularities, which can be arbitrary close to the physical region. In the following section, a dispersive representation is constructed, 
which resolves these problems.

\begin{figure}[t!]
\begin{center}
\includegraphics[width=0.42\textwidth]{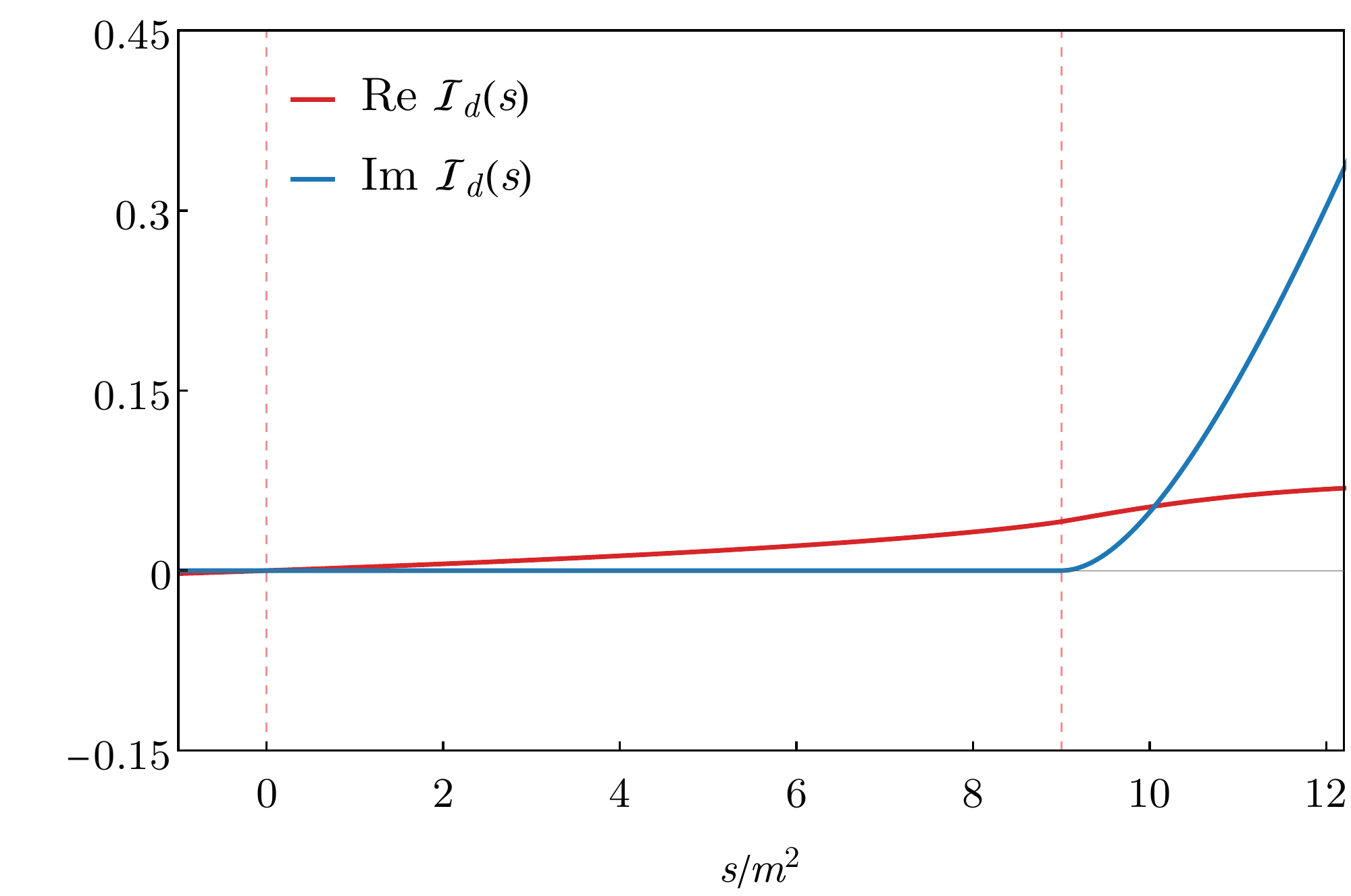}
\end{center}
\caption{The dispersed kernel $\Ic_d(s)$. The two-body bound state pole is at $M^2 = 3 m^2$ which corresponds to the energy $s_{\text{th},2} \approx 7.456 m^2$. Three-body threshold $s_{\text{th},3}$ is highlighted by the dashed, red line. Above  $s_{\text{th},3}$ the dispersed integral $\Ic_d(s)$ is identical to the original model $\Ic(s)$.}
\label{fig:dispersed-I}
\end{figure}

\subsection{The improved $B$-matrix formalism}

The problems with nonphysical singularities discussed in the previous section can be resolved by a dispersion representation. We use it to construct a new kernel $\Ic\to \Ic_d$ which inherits only the right-hand, three-body unitarity cut from $\Ic$, 
    \beq
    \label{eq:dispersed-I}
    \Ic_d(s) = \frac{(s-s_s)^2}{\pi}\!\!\!\! \int\limits_{(3m)^2}^\infty \!\!\! ds' \frac{\im \, \Ic(s')}{(s'-s - i \epsilon)(s'-s_s - i\epsilon)^2} ,~~~~
    \eeq
where $s_{s}$ is the subtraction point. In the following we set $s_s=0$. Two subtractions are needed given the asymptotic behavior of $\im \, \Ic(s)$. By construction, the imaginary part of $\Ic_d(s)$ is zero below $s_{\text{th},3}$. The plot of $\Ic_d(s)$ is shown in Fig.~\ref{fig:dispersed-I}. For completeness, we also replace the two-body phase space $i \rho_2(s) \to i\rho_{2,d}(s)$ by the Chew-Mandelstam function which removes the nonphysical singularity at $s=0$,
    \beq
    \label{eq:dispersed-rho}
    i \rho_{2,d}(s) = \frac{s}{\pi}\!\!\! \int\limits_{(M+m)^2}^\infty \!\!\!\!\! ds' \ \frac{  \rho_2(s')}{s'(s' - s )} \, .
    \eeq
Equivalently, to remove left-hand singularities from $\rho_2(s)$ one can also consider an ``improved" version
    \beq
    i\rho_{2,\text{imp}}(s) = - \frac{1}{16\pi} \sqrt{ (M+m)^2-s} \ .
    \eeq
Note that with these replacements, the amplitudes are now defined on the physical sheet.

\begin{figure}[t!]
    \centering
    \includegraphics[ width=0.3\textwidth,trim= 2 2 2 2,clip]{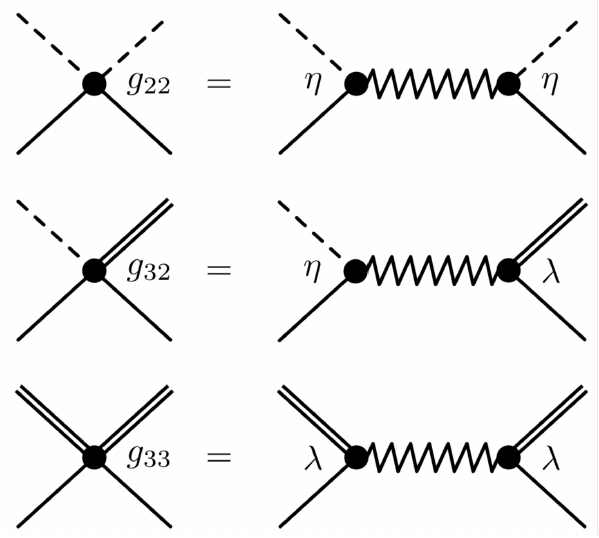}
    \caption{Motivation for the $G=0$ condition. In the microscopic model the contact interaction amplitudes $g_{mn}$ (small black circles) are explained as a heavy particle exchange (``zig-zag" line) which couples with a strength $\lambda$ to three particles and $\eta$ to a bound-state and a particle. This gives $g_{22} \sim \eta^2$, $g_{33} \sim \lambda^2$ and $g_{32} \sim \eta \lambda$, which satisfies $g_{32}^2 = g_{22} g_{33}$.}
    \label{fig:G=0}
\end{figure}

\begin{figure}[t!]
\begin{center}
\subfigure[~Dispersed kernel $\Ic_d(s_{\text{th},2})$ as a function of $a_0$.]
{
\includegraphics[width=0.43\textwidth]{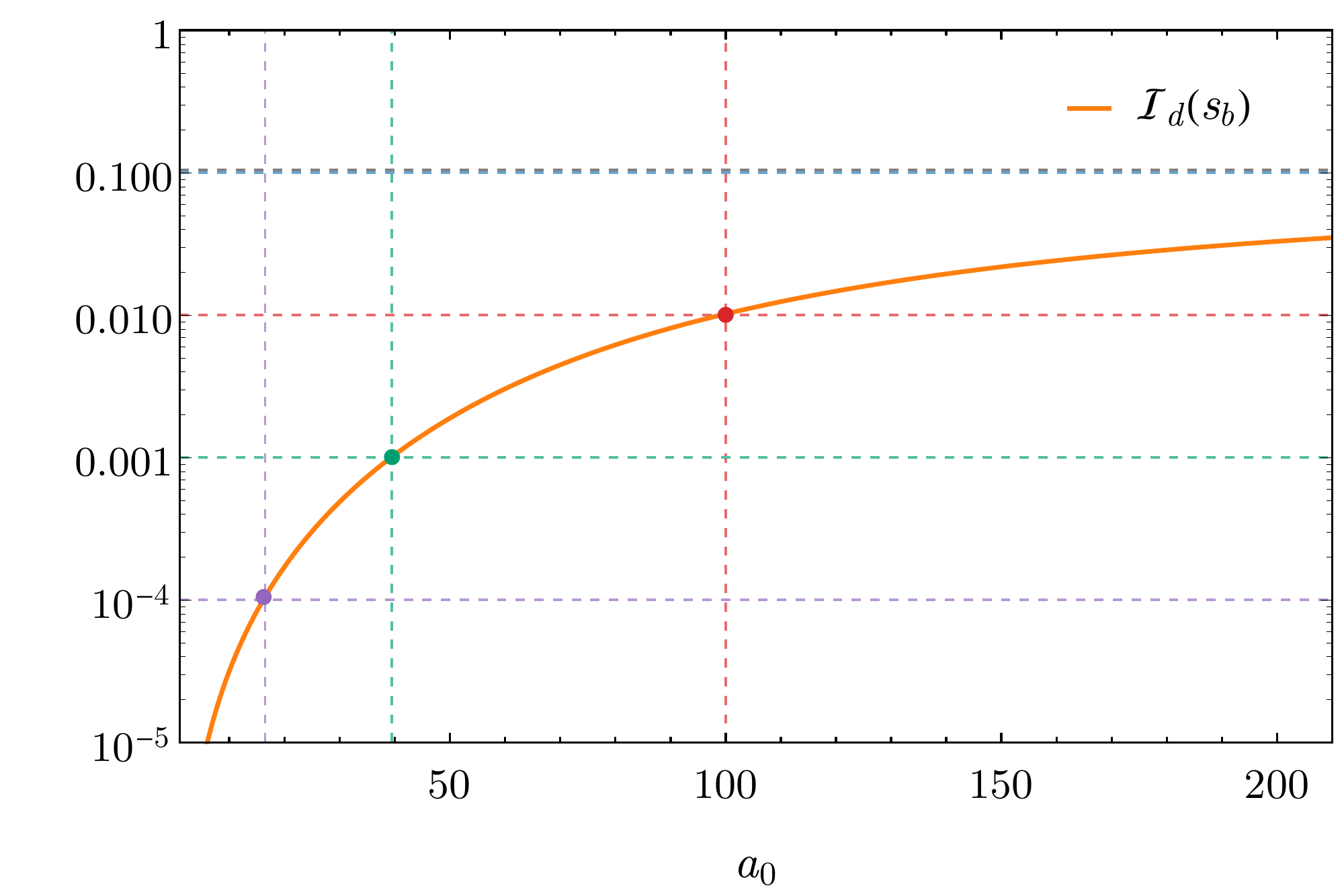}
\label{fig:b0-a}
}
\subfigure[~Bound-state--spectator scattering length $b_0$ vs. $a_0$.]
{
\includegraphics[width=0.42\textwidth]{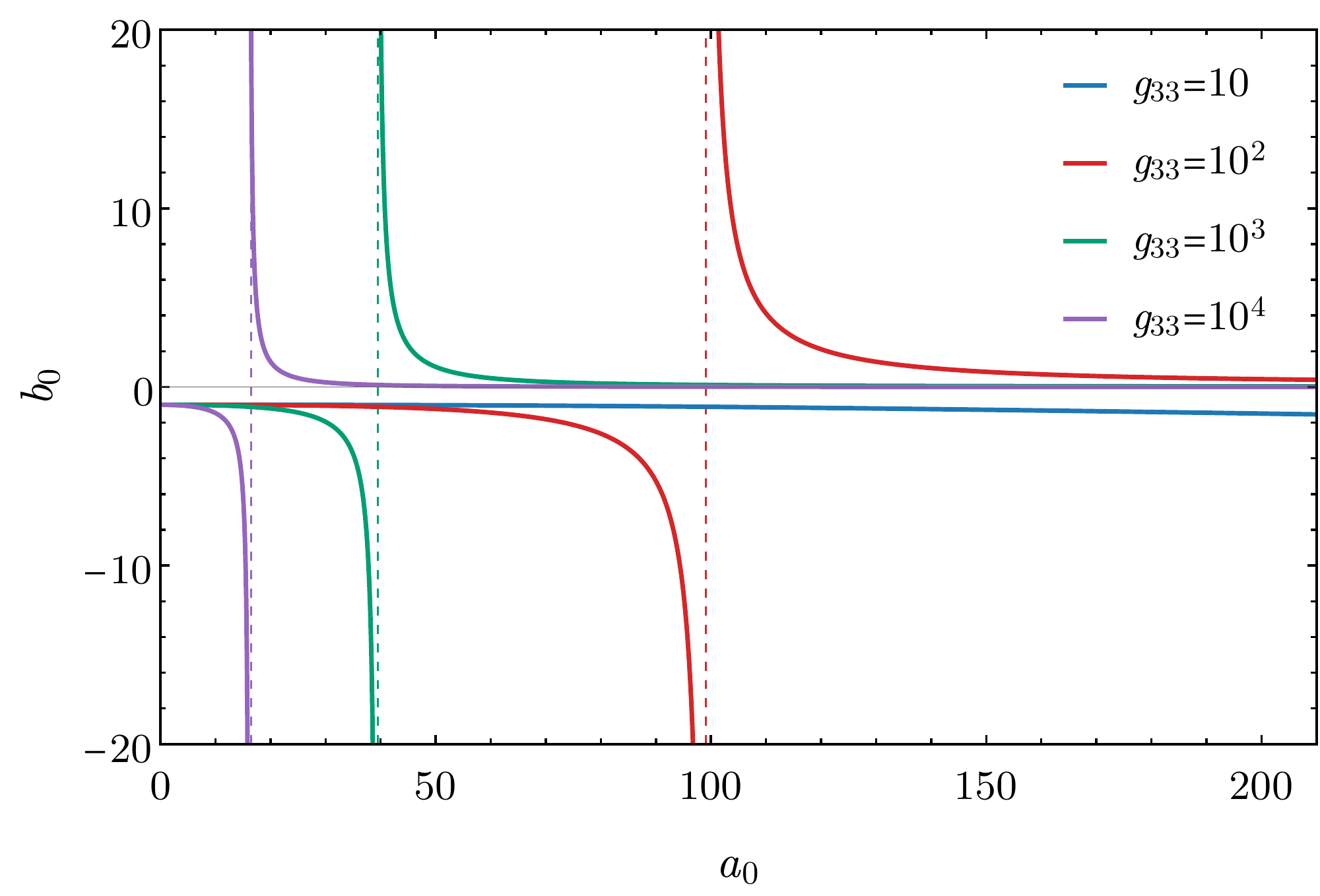}
\label{fig:b0-b}
}
\end{center}
\caption{ Panel (a): Dispersed kernel $\im \, \Ic_d(s_{\text{th},2})$ as a function of $a_0$ on the logarithmic plot. Color lines correspond to the inverted couplings of the Fig.~\ref{fig:b0-b}, while color points are solutions of the $\Ic_d(s_{\text{th},2}) = 1/g_{33}$ three-body bound-state equation. Panel (b): Dependence of the bound-state--spectator scattering length $b_0$, defined in Eq.~\eqref{eq:KBmatrix3} for $G=0$, on the two-body dimensionless scattering length $a_0$. Results for various values of $g_{33}$ are shown, while the coupling $g_{22}=1$. It can be seen that for large enough $g_{33}$ and particular values of $a_0$ a shallow three-body bound state is developed, with a mass equal to $M+m$.}
\label{fig:fig:b0_vs_a0}
\end{figure}

The improved (or ``dispersed") $B$-matrix parametrization results in the $\2\to\2$ amplitude with the proper analytical behavior below $s_{\text{th},3}$, 
    \beq
    \label{eq:d22solB}
    \widetilde{a}_{22,d}(s) &=& \frac{1}{ \frac{1}{\Kc(s)} - i \rho_{2,\text{imp}}(s) } \, ,
    \eeq
where the real $K$ matrix is given by,
    \beq
    \label{eq:KBmatrix}
    \Kc(s) = \frac{g_{22} + G \, \Ic_d(s)}{1-g_{33} \, \Ic_d(s) } \, .
    \eeq
The above general formula can be simplified by choosing $G=0$, i.e. for couplings to factorize, $g_{22} g_{33} = g_{23}^2$. This would happen for example in a microscopic model in which $g_{mn}$ are effective couplings originating from an exchange of a heavy mediator, see~Fig.~\ref{fig:G=0}. In this case, at $s = s_{\text{th},2}$, for which the relative momentum between the bound-state and spectator  vanishes, 
$\Kc$ becomes the bound-state--particle scattering length $b_0$. It is determined by both the two-body and three-body interactions,
    \beq
    \label{eq:KBmatrix3}
    b_0 = -\frac{g_{22}}{\Big( 1 - g_{33} \, \Ic_d(s_{\text{th},2})\Big)} \ .
   \eeq
We see that for $\Ic_d(s_{\text{th},2}) = 1/g_{33}$ the scattering length $|b_0| \to \infty$ and a shallow three-particle bound state appears at the
$s_{\text{th},2}$ threshold, see Fig.~\ref{fig:fig:b0_vs_a0}. As expected, this is a purely three-body effect, as it does not depend on the value of the bound-state--spectator coupling $g_{22}$, which is responsible for scaling of $b_0$. Since $\Ic_d(s_{\text{th},2})$ is a monotonic function of $a_0$, there can exist only one three-body bound state for a given two-body scattering length. The $\3 \to \3$ improved $B$-matrix amplitude becomes:
    \beq
    \label{eq:d33solB}
    \widetilde{a}_{33,d}(s) &=& \frac{1}{\frac{1 - g_{22} i \rho_{2,\text{imp}}(s)}{g_{33} + G i \rho_{2,\text{imp}}(s)} - \Ic_d(s) } \nln
    &\stackrel{G\to0}{=}& \left(\frac{g_{33} }{ g_{22} } \right) \frac{1}{\frac{1}{\Kc(s)}
    - i \rho_{2,\text{imp}}(s)  }  \ ,
    \eeq
and therefore below the three-particle threshold,
    \beq
    \label{eq:d33solB2}
    \im [\widetilde{a}_{33,d}(s) ]^{-1} &=& - \rho_{2,\text{imp}}(s) \left[ \frac{ g_{32}^2 }{g_{33}^2 + G^2 [\rho_{2,\text{imp}}(s)]^2} \right]~~~~ \nln
    &=& - \rho_{2,\text{imp}}(s) \left| \frac{\widetilde{a}_{32,d}(s)}{\widetilde{a}_{33,d}(s)} \right|^2 .
    \eeq
Which now agrees with the unitarity relation in Eq.~\eqref{eq:unitarity-check33}.

Finally, it is interesting to compare the above results with the EFT-like model of Eq.~\eqref{eq:d33solEFT2}. The EFT-like model $\2\to\2$ and $\3\to\3$ amplitudes are related by the proportionality factor of $g^2$. In the improved $B$-matrix model, the relation between $\widetilde{a}_{22} $ and $\widetilde{a}_{33}$ is $s$-dependent. This is to be expected in general, when there are no constraints between two- and three-body interactions. When factorization of couplings is imposed, the amplitudes become proportional. By requiring that the improved $B$-matrix amplitudes are related in exactly the same way as in the EFT-like model one can establish a relation between the couplings and the two-body residue $g$,
    \beq
    \label{eq:couplings-relation}
     g^2 = \left( \frac{g_{32}}{g_{33}} \right)^2 = \frac{g_{22} }{g_{33} } \, .
    \eeq
Moreover, to ensure that results of the dispersed $B$-matrix model and the EFT-like model agree with each other, it is required they have identical scattering lengths $b_0$. This condition allows us to relate the three-body EFT-like coupling to the $B$-matrix coupling, namely:
    \beq
    \label{eq:couplings-relation2}
    \frac{1}{g_{33,\text{EFT}}} + \frac{1}{g_{33}} = \Ic_d(s_{\text{th},2}) \, .
    \eeq
To conclude, for fixed fixed $g_{33}$ and $a_0$, condition $G=0$ together with Eqs.~\eqref{eq:couplings-relation} and \eqref{eq:couplings-relation2}, allows us to reproduce the results of the EFT-like model from the dispersed $B$-matrix formalism and vice versa.

\section{Conclusions}
\label{sec:conclusions}

In this work, the analytic features of the $B$-matrix formulation were discussed. We emphasized that within this framework intermediate particles are on-shell. This allows for a clear interpretation of the long-range interactions, by identifying them with an exchange of physical particles. However, it leads to spurious left-hand singularities and prevents one from extracting amplitudes involving a two-body bound-state from the three-body ones.

We have shown how these shortcomings are eliminated when the physical, coupled channels are included and dispersion relations are used to push the spurious singularities into nonphysical sheets. A generalization of the elastic $\3\to\3$ $B$-matrix was proposed, i.e. a parametrization which includes $\2\to\2$, $\2\to\3$ and $\3\to\2$ scattering channels explicitly. It was shown that above the three-body threshold the new $B$-matrix equations satisfy the unitarity relation for the multi-channel $S$-matrix. We introduced a dispersion procedure that removed artificial left-hand branch points from the integration kernel and led to controllable and correct amplitudes, which satisfy unitarity constraint above all relevant thresholds. Our analysis provides an additional argument for the necessity of implementing a dispersion procedure for the $B$-matrix kernel, which was first shown in the study of the triangle diagram in Ref.~\citep{Jackura:2018}.

Recent results concerning the spectrum of the $\pi\pi\pi$ system on the lattice \citep{Mai:2018, Mai:2019, Horz:2019, Culver:2019vvu, Blanton:2019, Sadasivan:2020, Fischer:2020, Hansen:2020otl} open a possibility of applying the $B$-matrix to the real physical problems, such as the analysis of the $X(3872)$ resonance, which will be explored in the future.


\begin{acknowledgements}
We thank Andrew Jackura for many useful discussions and for the permission to use some of his elegant diagrams in this article. This work was supported by the U.S. Department of Energy under Grants No. DE-AC05-06OR23177 and No. DE-FG02-87ER40365.
\end{acknowledgements}

\appendix

\section{Conventions and kinematics for the three-body elastic scattering}
\label{app:A}

This appendix is based on the introductory discussion of Ref.~\citep{Jackura:2019}. It is intended as a short review of the notation, definitions, and relevant three-body kinematic quantities. Additional details can be found in Sec.~2 and App.~A of Ref.~\citep{Jackura:2018} or in textbooks~\citep{Byckling:1971}. 

\subsection{Three-body kinematics}

We consider an elastic scattering process, in which incoming and outgoing state consists of three indistinguishable particles of mass $m$. The initial total four-momentum is denoted $P=(E,\P)$ and the final $P'=(E',\P')$, with the total invariant mass squared of the system being $s = P^2 = P'^2$. The three particles in the final and initial states are divided into an isobar (a pair) and a spectator (a single particle). The initial spectator has a four-momentum $p=(\omega_{\p}, \p)$, with the on-shell energy being $\omega_{\p} = \sqrt{\p^2+m^2}$, while the corresponding initial isobar is characterized by four-momentum $P_{\p} = (E_{\p}, \P - \p) = (E -\omega_{\p}, \P - \p)$ and its invariant mass squared $\sigma_{\p} = P_{\p}^2$. For the on-shell isobar it belongs to the interval $\sigma_{\p} \in [4m^2 , (\sqrt{s} - m)^2]$. The magnitude of the spectator momentum can be related to the corresponding isobar invariant mass squared via formula 
    \beq
    |\p| = \frac{1}{2\sqrt{s}} \, \lambda^{1/2}(s,m^2,\sigma_{\p}) \, ,
    \eeq
where 
    \beq
    \lambda(x,y,z) = x^2 + y^2 + z^2 - 2xy - 2yz - 2zx
    \eeq
is the K\"allen triangle function. Four-momenta of two constituents of the isobar are $q_{\p}$ and $P - p - q_{\p}$, respectively. Finally, analogous variables for outgoing particles are denoted with a prime, e.g. the outgoing spectator's four-momentum is $p'=(\omega_{\p'},\p')$. 

\begin{figure}[t!]
    \centering
    \includegraphics[ width=0.8\columnwidth]{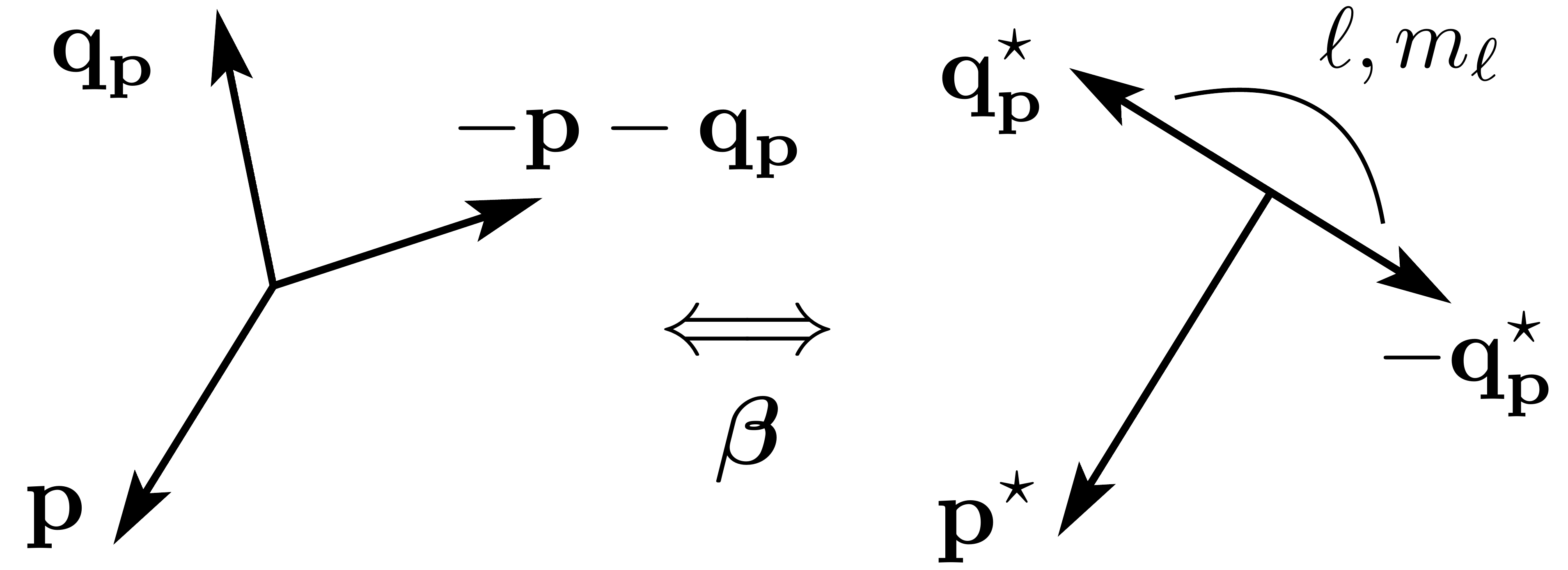}
    \put(-170,0){\colorbox{white}{(a)}}
    \put(-50,0){\colorbox{white}{(b)}}
    \caption{A three-particle state in the (a) CMF and (b) HF. Lorentz boost with $\bs{\beta} = -(\P - \p) / (E - \omega_{\p})$ transforms between the frames. The angular momentum of the pair $(\ell,m_\ell)$ is defined in the HF with respect to the spherical angles of $\q_{\p}^\star$.}
    \label{fig:diag_3to3_CMF_boost}
\end{figure}

In this paper, the center-of-mass frame (CMF) is assumed, i.e. total momenta $\P=\P'=0$. However, one also distinguishes the rest frame of the isobar, called the helicity frame (HF)\footnote{In Ref.~\citep{Jackura:2018} it is just called the isobar rest frame (IRF).}, in which a notion of the isobar angular momentum can be introduced. The HF is denoted by a ``$\star$" symbol and defined by a condition $\P^{\star} - \p^{\star} = \0$. In this frame the particles inside the pair have momenta $\q_{\p}^{\star}$ and $-\q_{\p}^{\star}$ while the spectator momentum $\p^\star$ makes the $-z$ axis.
The relative momentum of the pair in the HF can be expressed by the $\sigma_{\p}$ invariant and is equal to
    \beq
    |\q_{\p}^{\star}| = \frac{1}{2}\sqrt{\sigma_{\p} - 4m^2} \, .
    \eeq
Since the initial and final isobars are different, so are their associated helicity frames. It is useful to write down formulas for magnitudes of the momentum of the initial spectator in the final HF, $\p_{\p'}^\star$, and of the final spectator in the initial HF, ${\p_{\p}'}^\star$, in terms of the CMF quantities. They are
    \begin{align}
    \label{eq:spectator_cross}
    |\p_{\p'}^{\star}| & = \frac{1}{2\sqrt{\sigma_{\p'}}} \, \lambda^{1/2}\left((P_{\p'} - p )^2,\sigma_{\p'},m^2 \right) \, , \\
    |\p_{\p}'^{\star}| & = \frac{1}{2\sqrt{\sigma_{\p}}} \, \lambda^{1/2}\left((P_{\p} - p' )^2,\sigma_{\p},m^2 \right) \, ,
    \end{align}
From the energy-momentum conservation one has $P_{\p} - p' = P_{\p'} - p$. The kinematics of three particles is illustrated in Fig.~\ref{fig:diag_3to3_CMF_boost}.

\subsection{Three-body amplitude}

The elastic $\3\to\3$ scattering amplitude, $\Mc_{33}$, is defined as:
    \beq
    \label{eq:Smatrix}
    \bra \textrm{out} | T_{33} | \textrm{in} \ket  = (2\pi)^{4} \delta^{(4)}(P' - P) \,  \Mc_{33},
    \eeq
where the $T_{33}$-matrix is given by $S = \mathbbm{1} + i \, T_{33}$. Here the ``$33$" subscript is used, absent in Ref.~\citep{Jackura:2019}, to distinguish $\Mc_{33}$ from other amplitudes appearing in the more general treatment of Sec.~\ref{sec:multi-channel}. Amplitude $\Mc_{33}$ is symmetrized with respect to exchanges within any pair of incoming (outgoing) particles, however, it is easier to work with the unsymmetrized amplitude $[\,\Mc_{\p'\p}\,]_{\ell' m_{\ell}' ; \ell m_{\ell} }$, written in the so-called $(\p \ell m_{\ell})$ basis. It can be treated as an infinite-dimensional matrix in the angular momentum space. The amplitude depends on eight variables: initial and final isobar invariant masses squared $\sigma_{\p}$, $\sigma_{\p'}$, total invariant mass squared of three particles $s$, angle between incoming and outgoing spectator $\Theta_{\p'\p}$ (or equivalently the total angular momentum $J$, see the discussion in Ref.~\citep{Jackura:2018}), and angular momenta of isobars $(\ell, m_\ell)$ and $(\ell', m'_\ell)$. The symmetric amplitude is related to the unsymmetrized one by
    \beq
    \label{eq:SymmAmp}
    \Mc_{33} \! = \! \text{Sym} \! \! \left[ \! 4\pi \!\! \sum_{ \substack{ \ell',m_{\ell}' \\ \ell,m_{\ell}} } \! Y_{\ell' m_{\ell}'}\!({\bh{\q}}_{\p'}^{\star}) \!
    \left[ \Mc_{33,\p'\p} \right]_{\ell' m_{\ell}' ; \ell m_{\ell}} 
    \! Y_{\ell m_{\ell}}^{*}\!({\bh{\q}}_{\p}^{\star}) \!\right] \!\! , \nonumber \\
    ~
    \eeq
where the operation $\text{Sym}[\bullet]$ means symmetrizing with respect to particle permutations. We omit the angular momentum labels when convenient and simplify the $\sigma_{\p}$ and $s$ dependence notation by using just the spectator momentum subscripts, i.e. by writing $\Mc_{33,\p'\p}$. The $\p$-dependence will be left explicit unless otherwise noted.

The amplitude $\Mc_{33,\p'\p}$ is be separated into a connected and disconnected parts, $\Ac_{\p'\p}$ and $\Fc_{\p}\, \delta_{\p'\p}$, respectively,
    \beq
    \label{eq:dis-con-decomp}
    \Mc_{33,\p'\p} = \Ac_{\p'\p} + \Fc_{\p} \, \delta_{\p'\p} \, .
    \eeq
In the disconnected piece, the momentum conserving $\delta$-function is included, $\delta_{\p'\p} = (2\pi)^{3} \, 2\omega_{\p} \, \delta^{(3)}(\p'-\p)$. The disconnected part has a form of a two-body scattering amplitude in the isobar sub-channel, with the spectator not taking part in the interaction. It depends on the isobar angular momentum and its invariant mass squared $\sigma_{\p}$. It is represented by the diagram:
    \begin{align}
    \left[ \, \Fc_{\p} \, \right]_{\ell' m_{\ell}' ; \ell m_{\ell}}  & = \delta_{\ell' \ell} \, \delta_{m_{\ell}' m_{\ell}} \, \Fc_{\ell} (\sigma_{\p}) \nonumber \\
    & = \includegraphics[width=0.25\textwidth, valign=c]{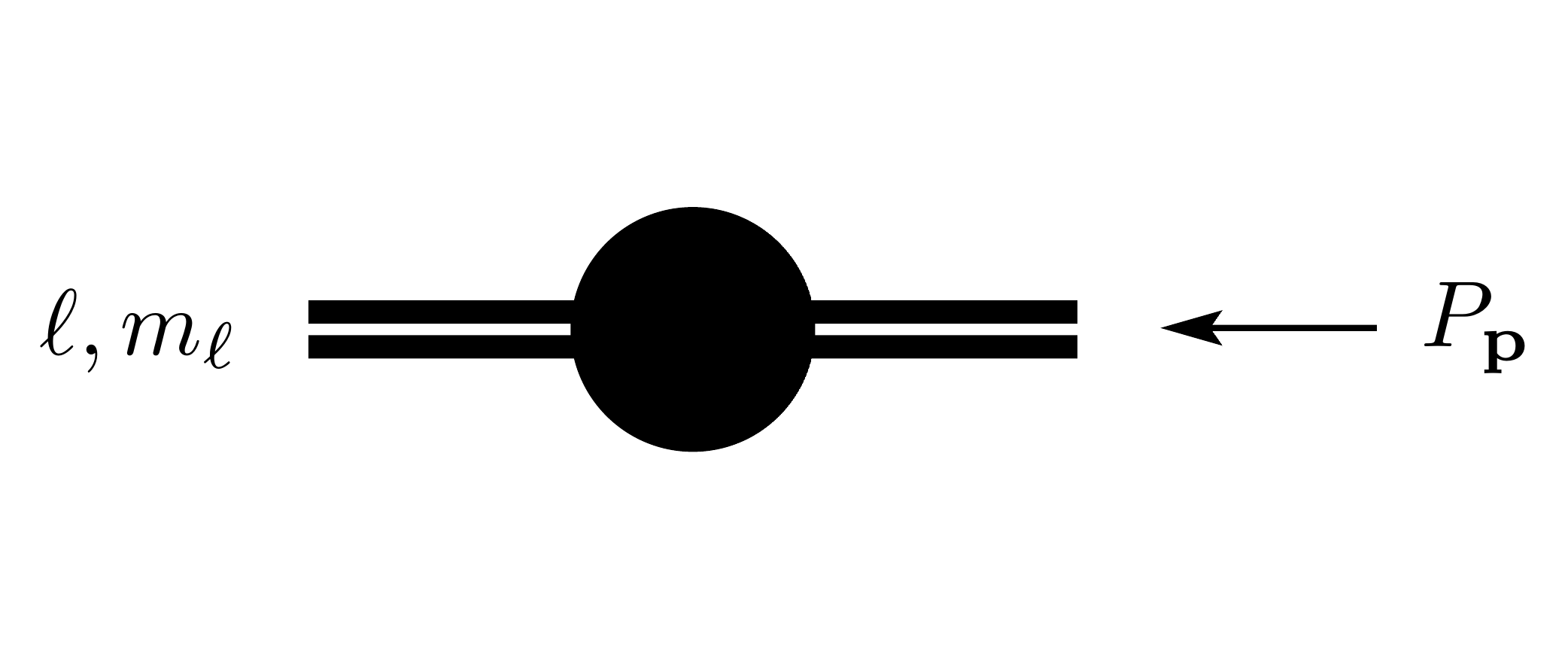} 
    \end{align}
The connected term $\Ac_{\p'\p}$ contains off-diagonal contributions in spin indices. The corresponding diagram is:
    \begin{align}
    \left[ \, \Ac_{\p'\p} \, \right]_{\ell' m_{\ell}' ; \ell m_{\ell}} & = \Ac_{\ell' m_{\ell}' ; \ell m_{\ell}} (\p',s,\p) \nonumber \\
    & = \includegraphics[width=0.25\textwidth, valign=c]{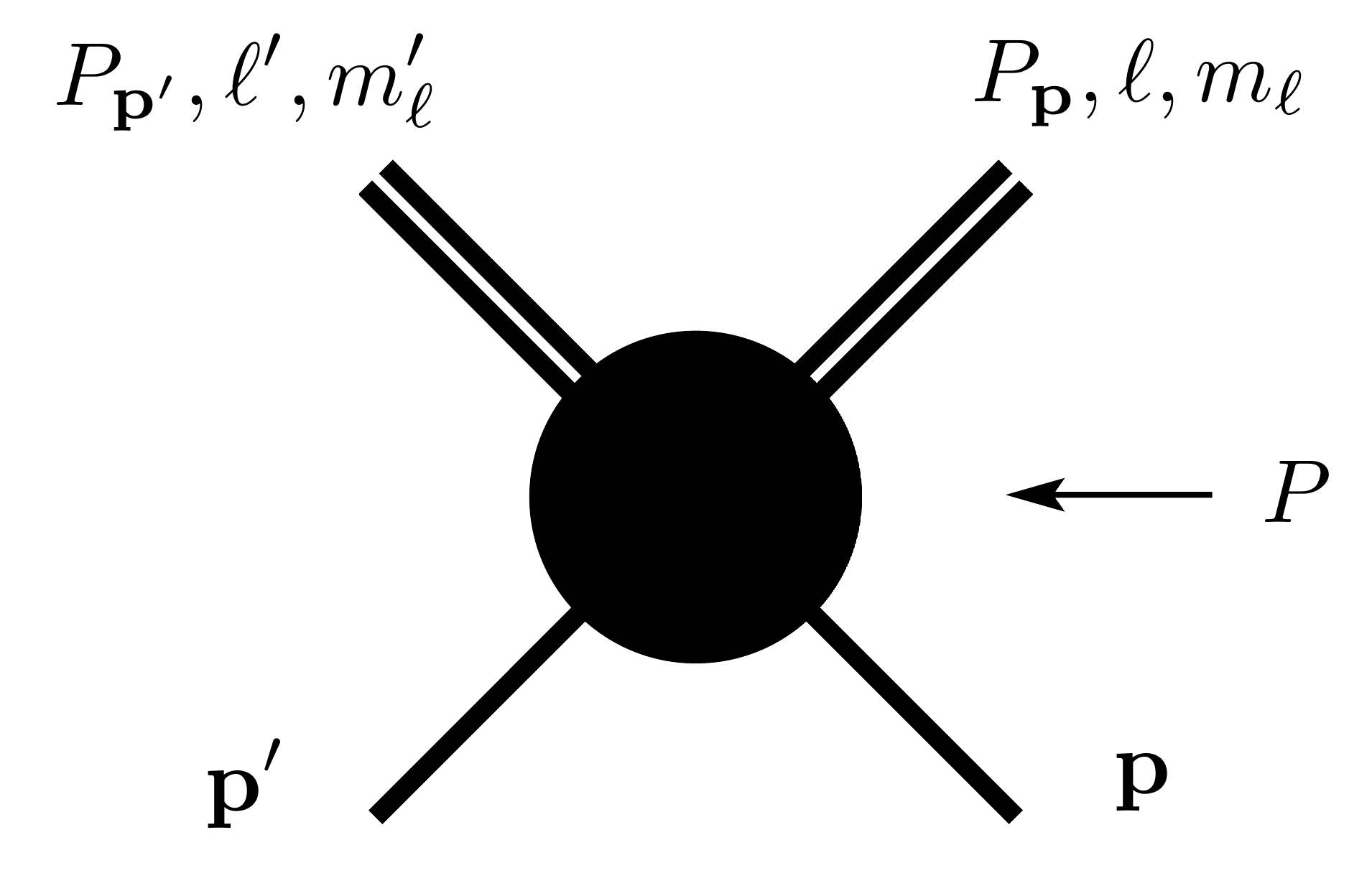}
    \end{align}

The $B$-matrix parametrization of Eq.~\eqref{eq:b-matrix-param} is built upon the integral kernel $\Bc_{33}$. It contains the OPE amplitude $\Gc_{\p'\p}$ and a real function $\Rc_{\p'\p}$ called the $R$-matrix, see Eq.~\eqref{eq:B-matrix-decomp}. The OPE amplitude is
    \begin{align}
    \label{eq:OPE}
    \left[ \,\Gc_{\p'\p} \, \right]_{\ell' m_{\ell}' ; \ell m_{\ell}} & = \left( \frac{p_{\p'}^{\star}}{q_{\p'}^{\star}} \right)^{\!\!\ell'}\!\! \frac{4\pi \, Y_{\ell' m_{\ell}'}^{*}({\bh{\p}}_{\p'}^{\star}) Y_{\ell m_{\ell}}({\bh{\p}}_{\p}'^{\star}) }{m^2 - (P_{\p} - p')^2 - i\epsilon} \left( \frac{p_{\p}'^{\star}}{q_{\p}^{\star}} \right)^{\!\!\ell} \nonumber \\
    & = \includegraphics[width=0.25\textwidth, valign=c]{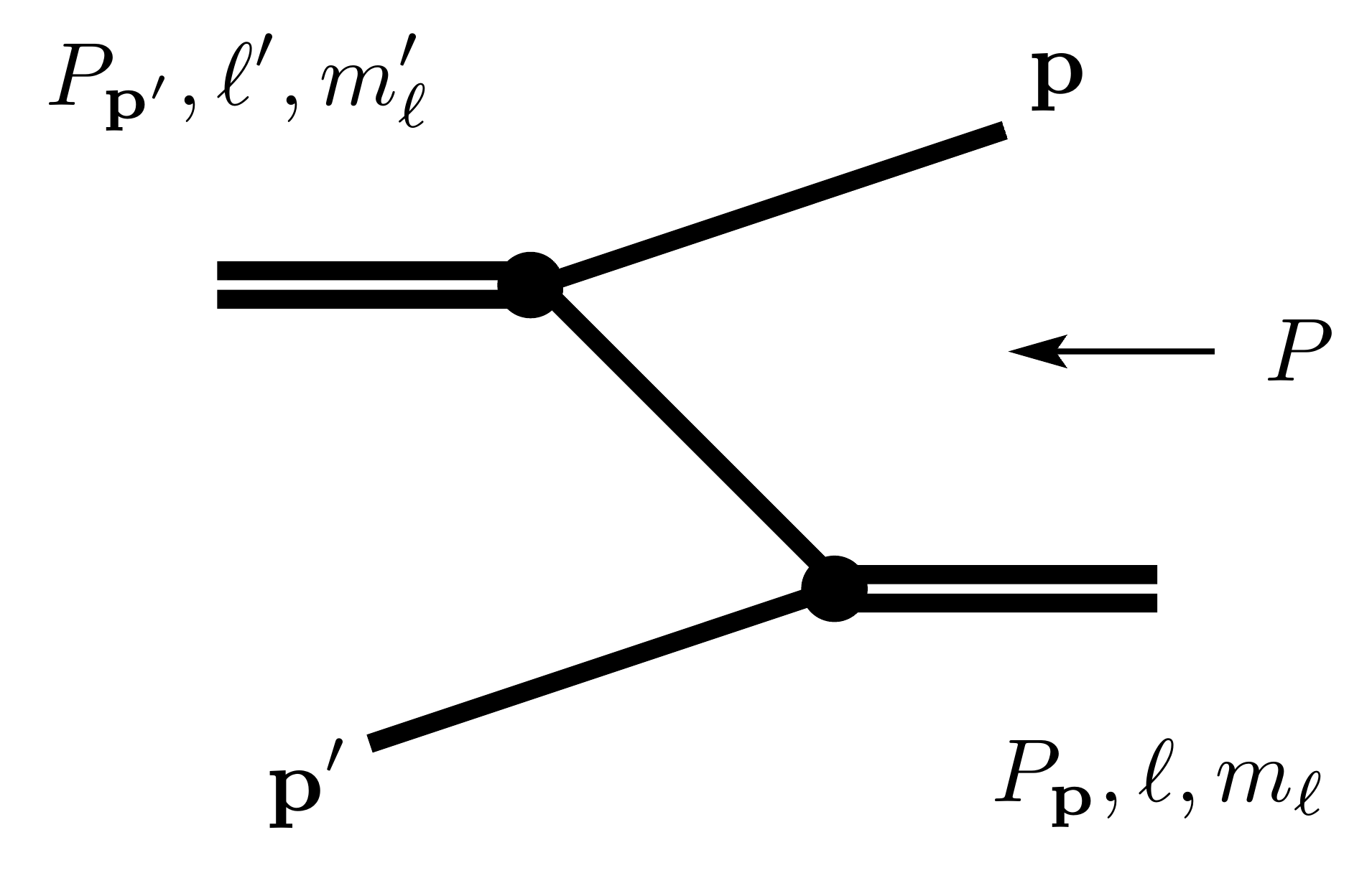}
    \end{align}
and describes the probability of a particle exchange between the isobar and a spectator. The form of the $R$-matrix is unconstrained by the unitarity principle.

\section{The NREFT ladder equation}
\label{app:B}

In Sec.~V of Ref.~\citep{Jackura:2019}, it was shown that the non-relativistic limit of Eq.~\eqref{eq:b-matrix-param} reproduces properly the Faddeev equations of Ref.~\citep{Elster:2008}, showing that Eq.~\eqref{eq:b-matrix-param} can be used to describe the low-energy physics. Here we derive the correspondence between the non-relativistic EFT (NREFT) approach of Ref.~\citep{Bedaque:1998} and the $B$-matrix equations. There, a $S$-wave scattering of a dimer (two-body bound state) with a single scalar particle is considered, with the inclusion of the short-range, three-body forces. In the region of applicability, the predictions of NREFT agree numerically with an implementation of the relativistic three-body quantization condition in Ref.~\citep{Romero-Lopez:2019} and solution of the relativistic three-body ladder equation in Ref.~\citep{Jackura:2020bsk}. Here we focus on the analytic relation between the relativistic and non-relativistic versions of the ladder amplitude.

We consider the $S$-wave scattering, Eq.~\eqref{eq:ladder-s-wave}, and include the momentum-independent $R$-matrix $\Rc_{\p'\p} = r$. One can introduce the non-relativistic kinetic energy of three particles $s= ( \Delta E + 3m )^2$ and expand $\omega_{\p} = m + \p^2/2m + \Oc(\p^4)$. Here, the non-relativistic notation $p = |\p|$, $p'=|\p'|$, $q=|\q|$ is used. The argument of the OPE reduces to
    \beq
    z_{\p\p'} \approx \frac{1}{p p'} \left( m \Delta E - p^2 - p'^2 \right) \, ,
    \eeq
therefore
    \beq
    \Gc_{\text{nr}}(p',\Delta E,p)\! &=& \!\frac{1}{4 p' p} \times \nln
    & & \log \left(\!\frac{ m \Delta E - p'^2 - p^2 - p p'}{ m \Delta E - p'^2 - p^2 + p p'}\! \right) \, . ~~~
    \eeq
From Eq.~\eqref{eq:2to2scatt} one derives the low-energy isobar amplitude
    \beq
    \label{eq:2to2amp2-non-rel}
    \Fc^{-1}_{\text{nr}}(q) = \left( - \frac{32 \pi m }{a_0} + \sqrt{\frac{3q^2 }{4} - m \Delta E } \right) \frac{1}{32\pi m } \, .~~
    \eeq
Moreover converting back to the momentum integral via Eq.~\eqref{eq:integration} one obtain
    \beq
    \label{eq:meaasure}
    \int \frac{d \sigma_{\q}}{2\pi} \frac{\lambda^{1/2}(s,m^2,\sigma_{\q})}{8\pi s} = \int \frac{d q \ q^2}{2\pi^2 2 \omega_{\q}} \, .
    \eeq
The overall factor from Eq.~\eqref{eq:2to2amp2-non-rel} and integration measure in Eq.~\eqref{eq:meaasure} simplifies in the non-relativistic limit to
    \beq
    \frac{32 \pi m}{2 \pi^2 2 \omega_{\q}} = \frac{4\sqrt{E^2 - 2 E \omega_{\q} + m^2 }}{\pi \sqrt{q^2 + m^2}} \approx \frac{8}{\pi} \, .
    \eeq
Finally, one defines a non-relativistic scattering length of the proper dimension, $32 \pi m/a_0 = 1/a_2$, and obtains a non relativistic equation for the three-body connected amplitude $a_{33,\text{nr}}(p', p)$,
    \beq
    \label{eq:ladder1}
    & & a_{33,\text{nr}}(p', p) = \Gc_{\text{nr}}(p',\Delta E,p) + r \nln
    & & + \frac{8}{\pi} \!\!\! \int\limits_0^{q_{\text{max,nr}}} \!\! dq \, q^2  \frac{a_{33,\text{nr}}(q,\Delta E, p)}{ - \frac{1}{a_2} + \sqrt{ \frac{3q^2}{4} -m\Delta E} }
     \big[\Gc_{\text{nr}}(p',\Delta E,p)  + r \big] \, .
    \eeq
The integration interval $[4m^2, (\sqrt{s}-m)^2]$ is transformed into the corresponding non-relativistic interval $[0, q_{\text{max,nr}} = \sqrt{4m \Delta E/3 }]$. This result can be compared with Eq. (7) in Ref. \citep{Bedaque:1998},
    \beq
    \label{eq:ladder-Hammer}
    t(k,p) &=& \frac{mg^2}{pk} \log \left( \frac{p^2 + pk +k^2 -mE}{p^2 -pk +k^2 -mE} \right) + h \nln 
    &+& \frac{2}{\pi} \int_0^\Lambda dq \  q^2 \ \frac{t(k,q) }{-\frac{1}{a_2} + \sqrt{\frac{3q^2}{4} -mE } } \nln
    & & \times \left[ \frac{1}{pq} \log \left( \frac{p^2 + pq +q^2 -mE }{p^2 -pq +q^2 -mE} \right) + \frac{h}{mg^2} \right] \, ,
    \eeq
where $E = \frac{3k^2}{4m} - B_2$ and $B_2$ is the binding energy of the bound state. The constant $g$ is a coupling of the effective field theory and $h$ describes strength of the three-body contact interactions. Both equations look similar, up to the difference in normalization
    \beq
    a_{33,\text{nr}}(p',\Delta E, p) = \frac{t(p',p)}{4mg^2} ,
    \eeq
the value of energy (since the three-particles energy has not yet continued to the bound state energy $\Delta E \to \Delta E_b = \frac{3p^2}{4m} - \frac{1}{m a_2^2}$), but most importantly, the choice of the regularization parameter. Because for the bound state energy $q_{\text{max,nr}} = \sqrt{3m\Delta E_b/3}$ the bound-state pole is not included in the integration interval and $a_{33,\text{nr}}$ is not a correct bound-state--spectator scattering amplitude.

Indeed, in Ref.~\citep{Bedaque:1998} an off-shell two-body amplitude $a(p',p)$ is defined
    \beq
    \frac{a_t(p',p)}{p^2-p'^2 + i \epsilon} = \frac{1}{mg^2} \frac{t(p',p)}{ - \frac{1}{a_2} + \sqrt{ \frac{3p'^2}{4} -m\Delta E} } \ ,
    \eeq
which satisfies the Lippmann--Schwinger-like equation
    \beq
    a(k,p) &=& M(k,p) \nln
    & & + \frac{2}{\pi} \int_0^\Lambda dq \ M(q,p) \frac{q^2 }{q^2 - k^2 - i \epsilon} a(k,q) \, .
    \eeq
The kernel
    \beq
    M(q,p) &=& \frac{4}{3} \left(\frac{1}{a_2} + \sqrt{\frac{3p^2}{4}-m E} \right) \times \nln
    & & \left[ \frac{1}{pq} \log \left( \frac{p^2 + pq +q^2 -mE }{p^2 -pq +q^2 -mE} + \frac{h}{mg^2}\right) \right] \, . 
    \eeq
By splitting the ``propagator"
    \beq
    \frac{1}{q^2 - k^2 - i \epsilon} = i \pi \delta(q^2 - k^2) + \text{P.V}\left( \frac{1}{q^2 - k^2} \right) \, ,~~~~
    \eeq
for large $\Lambda$ one obtains
    \beq
    a(k,p) &=& M(k,p) + i M(k,p) a(k,k) k \nln
    & & + \frac{2}{\pi} \text{P.V} \int_0^\Lambda dq \ M(q,p) \frac{q^2 }{q^2 - k^2 } a(k,q) ,
    \eeq
and can express the amplitude in the form which explicitly satisfies the two-body unitarity,
    \beq
    a(k,k) &=& \frac{1}{\frac{1}{K(k,k)} - i k } \, ,
    \eeq
where $K(k,p)$ is a real function satisfying equation:
    \beq
    K(k,p) &=& M(k,p) \nln
    & & + \frac{2}{\pi} \, \text{P.V} \int_0^\Lambda dq \ M(q,p) \frac{q^2 }{q^2 - k^2 } K(k,q) .
    \eeq
However, with our choice of $q_{\text{max,nr}}$, the Dirac delta does not contribute to the momentum integral, and an incorrect imaginary part of $[a(k,k)]^{-1}$ below the three-body threshold.

\section{Unitarity of the multi-channel formalism}
\label{app:C}

In this appendix the results of Ref.~\citep{Jackura:2018} and App.~A of Ref.~\citep{Jackura:2019} are extended. First, conventions of the generalized multi-channel formulation are explained. Then follows the presentation of the unitarity relations for the amplitudes describing scattering in the coupled two- and three-particle channels. Finally, it is shown that the representation given in Eqs.~\eqref{eq:b-matrix-22}--\eqref{eq:b-matrix-33} satisfies those constraints.4

\subsection{Unitarity relations}

We consider a $\2\to\2$, $\3\to\2$, $\2\to\3$ and $\3\to\3$ coupled scattering processes with the bound state of mass $M$ in the two-body channel. The scattering matrix is decomposed as $S = \one + i T$ and the scattering amplitude for the $\n \to \m$ process $\Mc_{mn}$ is defined as a matrix element:
    \beq
    \bra \bm{m}' | T | \bm{n} \ket = (2\pi)^4 \, \delta^{(4)}(P'-P) \, \Mc_{mn} \ ,
    \eeq
where $|\bm{n} \ket = | \p_1 \p_2 \dots \p_n \ket$ is an incoming state of $n$ particles, which inherits its normalization from one-particle states: $\bra \1' | \1 \ket \equiv \bra \p_1' | \p_1 \ket = (2\pi)^3 \, 2 \omega_{\p} \, \delta^{(3)}(\p'_1-\p_1)$. From now on, the letter $\k$ is used exclusively to denote momenta of the two-particle states, and letter $\p$ for momenta of the three-particle states. The four-vectors $P,P'$ are incoming and outgoing total four-momenta of the particles. As described in App.~\ref{app:A} the amplitude $\Mc_{33}$ is symmetrized with respect to exchanges among incoming and outgoing particles and depends on eight kinematic variables: angular momenta and invariant masses of the incoming and outgoing isobars, total invariant mass, and an angle between incoming and outgoing spectator. The amplitude $\Mc_{23}$ (and $\Mc_{32}$) is symmetrized with respect to exchanges among three incoming (outgoing) particles and depends on five variables: incoming (outgoing) isobar angular momentum and invariant mass, total invariant mass and an angle between incoming and outgoing spectator. Finally, $\Mc_{22}$ does not need symmetrization---since in the two-body channel particles are not identical---and it depends on two variables: total invariant mass and an angle between the incoming and outgoing spectator. Similarly to the $\3\to\3$ case, for $\2\to\3$ and $\3\to\2$ amplitudes a particular choice of a spectator and a pair in the three-particle channel is made, and one can perform a partial wave expansion in the spherical angles of the pair relative momentum in its HF. Therefore, in the following, we use amplitudes, which are unsymmetrized in the three-body channels, generalizing the formulation of matrices in the $(\p\ell m_{\ell})$ basis from App.~\ref{app:A}. Namely, in this basis we define the multi-channel, unsymmetrized amplitudes as: a matrix $\Mc_{33,\p'\p}$, ``vectors" $\Mc_{23,\p}$ and $\Mc_{32,\p'}$, and a ``scalar" $\Mc_{22}$ .

The unitarity relation for the $S$-matrix $S^\dag S = \one$ is equivalent to $iT^\dag T = T-T^\dag$, which, after appropriate projections on the $|\2\ket$ and $|\3\ket$ scattering states subspaces can be expressed in terms of the scattering amplitudes $\Mc_{mn}$. They are
\begin{widetext}
    \beq
    \label{eq:unitarity-relation-22}
    \prod_{i=1}^2 \int_{\k''_i} (2\pi)^4 \delta^{(4)} \Big(\sum_{l} \k''_l-P \Big) \Mc^*_{22} \Mc_{22} 
    + \frac{1}{3!} \prod_{j=1}^3 \int_{\p_j''} (2\pi)^4 \delta^{(4)}\Big(\sum_{l} \p_l''-P \Big) \Mc_{23}^* \Mc_{32}  &=& 2 \im \Mc_{22} \, , \\
    \label{eq:unitarity-relation-32}
    \prod_{i=1}^2 \int_{\k''_i} (2\pi)^4 \delta^{(4)} \Big(\sum_{l} \k''_l-P \Big) \Mc^*_{32} \Mc_{22} 
    + \frac{1}{3!} \prod_{j=1}^3 \int_{\p_j''} (2\pi)^4 \delta^{(4)}\Big(\sum_{l} \p_l''-P \Big) \Mc_{33}^*\Mc_{32}  &=& 2 \im \Mc_{32} \ , \\
    \label{eq:unitarity-relation-23}
    \prod_{i=1}^2 \int_{\k''_i} (2\pi)^4 \delta^{(4)} \Big(\sum_{l} \k''_l-P \Big) \Mc^*_{22} \Mc_{23} 
    + \frac{1}{3!} \prod_{j=1}^3 \int_{\p_j''} (2\pi)^4 \delta^{(4)}\Big(\sum_{l} \p_l''-P \Big) \Mc_{23}^* \Mc_{33}  &=& 2 \im \Mc_{23} \ , \\
    \label{eq:unitarity-relation-33}
    \prod_{i=1}^2 \int_{\k''_i} (2\pi)^4 \delta^{(4)} \Big(\sum_{l} \k''_l-P \Big) \Mc^*_{32} \Mc_{23} 
    + \frac{1}{3!} \prod_{j=1}^3 \int_{\p_j''} (2\pi)^4 \delta^{(4)}\Big(\sum_{l} \p_l''-P \Big) \Mc_{33}^* \Mc_{33}  &=& 2 \im \Mc_{33} \ .
    \eeq
\end{widetext}
For clarity we do not write here the momentum dependence of the amplitudes explicitly---one should keep in mind that they depend on the incoming and outgoing momenta of scattered particles and, where integration is preformed, on the momenta of the intermediate states. Due to the existence of two channels, there are two types of integration involved: the first is over the momenta $\k''_1$ and $\k''_2$ of the intermediate on-shell bound-state and spectator, the second is over momenta $\p_1'',\p_2'',\p_3''$ of the three identical particles. The two-body phase space integral simplifies to
    \beq
    & & \prod_{i=1}^2 \int_{\k''_i} \! (2\pi)^4 \delta^{(4)} \Big(\sum_{l} \k''_l-P \Big)= \nln & & 2 \rho_2(s) \, \theta(s-s_{\text{th},2}) \int \! \frac{d \Omega_{\k}}{4\pi} \equiv 
    2\bar{\rho}_2(s) \int_{\hat{\k}} \, , ~~
    \eeq
with $\bar{\rho}_2(s) = \theta(s-s_{\text{th},2}) \rho_2(s) $ and the bound-state--particle threshold $s_{\text{th},2} = (M+m)^2$. The integral is performed over the spherical angle $\Omega_{\hat{\k}}$ of the relative momentum $\k$ between the intermediate bound state and spectator. The three-body integral is simplified to
    \beq
    & & \prod_{j=1}^3  \int_{\p_j''}  (2\pi)^4 \delta^{(4)}\! \Big(\sum_{l} \p_l''-P\Big) \nln
    & & = 2 \, \theta(s-s_{\text{sh},3}) \int_{\q}  \rho_{\q}  \int \frac{d\Omega_{\hat{\q}_{\p''}^\star}}{4\pi} \, , ~~~
    \eeq
in the case when no re-coupling between different intermediate pairs occurs. When it does, one writes instead
    \beq
    & & \prod_{j=1}^3 \int_{\p_j''} (2\pi)^4 \delta^{(4)}\Big(\sum_{l} \p_l''-P\Big) = \nln
    & & 2 \, \theta(s-s_{\text{sh},3}) \int_{\q'} \int_{\q} \, \pi \, \delta\left((P_{\q}-q')^2 + m^2\right)  \, , ~~
    \eeq
preserving unintegrated Dirac delta for the particle exchanged between pairs. The intermediate spectator has momentum $\p''$. Using the unsymmetrized amplitudes, and separating the $\3\to\3$ amplitude into connected and disconnected piece: $\Mc_{33,\p'\p} = \Fc_{\p} \delta_{\p'\p} + \Ac_{33,\p'\p}$ one arrives to the unitarity constraints
    \beq
    \nonumber
    \im \Ac_{22} &=& \bar{\rho}_2 \int_{\hat{\k}} \Ac_{22}^\dag \Ac_{22} + \int_{\q} \Ac_{23,\q}^\dag \rho_{\q} \Ac_{32,\q} \\
    &+&
    \int_{\q} \int_{\q'} \Ac_{23,\q}^\dag \Cc_{\q\q'} \Ac_{32,\q'} \, ,
    \label{eq:unitarity-22}
    \eeq

    \beq
    \nonumber
    \im \Ac_{32,\p'} &=& \bar{\rho}_2 \int_{\hat{\k}} \Ac_{32,\p'}^\dag \Ac_{22} 
    + \int_{\q} \Ac_{33,\p'\q}^\dag \rho_{\q} \Ac_{32,\q} \nln
    &+&
    \int_{\q} \int_{\q'} \Ac_{33,\p'\q}^\dag \Cc_{\q\q'} \Ac_{32,\q'}
    + \Fc_{\p'}^\dag \rho_{\p'} \Ac_{32,\p'} 
    \\
    &+& \int_{\q} \Fc_{\p'}^
    \dag \Cc_{\p'\q} \Ac_{32,\q} \ ,
    \label{eq:unitarity-32}
    \eeq

    \beq
    \nonumber
    \im \Ac_{23,\p} &=& \bar{\rho}_2 \int_{\hat{\k}} \Ac_{22}^\dag \Ac_{23,\p}
    + \int_{\q} \Ac_{23,\q}^\dag \rho_{\q} \Ac_{33,\q\p} \nln
    &+&
    \int_{\q} \int_{\q'} \Ac_{23,\q'}^\dag \Cc_{\q\q'} \Ac_{33,\q\p}
    + \Ac_{23,\p} ^\dag \rho_{\p} \Fc_{\p}
    \\
    &+& \int_{\q} \Ac_{23,\q}^\dag \Cc_{\p'\q} \Fc_{\p}  \ ,
    \label{eq:unitarity-23}
    \eeq

    \beq
    \nonumber
    \im \Ac_{33,\p'\p} &=& \bar{\rho}_2 \int_{\hat{\k}} \Ac_{32,\p'}^\dag \Ac_{23,\p}
    + \int_{\q} \Ac_{33,\p'\q}^\dag \rho_{\q} \Ac_{33,\q\p} \nln
    &+&
    \int_{\q} \int_{\q'} \Ac_{33,\p'\q'}^\dag \Cc_{\q\q'} \Ac_{33,\q\p}
    \nln
    &+& \int_{\q} \Fc_{\p'}^
    \dag \Cc_{\p'\q} \Ac_{33,\q\p}
    + \int_{\q} \Ac_{33,\p'\q}^\dag \Cc_{\p'\q} \Fc_{\p} \nln
    &+& \Fc_{\p'}^\dag \rho_{\p'} \Ac_{33,\p'\p}
    + \Ac_{33,\p'\p} ^\dag \rho_{\p} \Fc_{\p} \\
    &+& \Fc_{\p'} \Cc_{\p'\p} \Fc_{\p} \ .
    \label{eq:unitarity-33}
    \eeq
Here $\Cc_{\p'\p} = \im [\Gc_{\p'\p}]$, as in Ref.~\citep{Jackura:2019}, and the three-body-threshold step functions are included implicitly where necessary. Above equations can be considered a generalization of Eq.~(A4) ibid.

\subsection{Proof}

In order to prove that the multi-channel $B$-matrix representation of Eqs.~\eqref{eq:b-matrix-22}--\eqref{eq:b-matrix-33} satisfies unitarity relations in Eqs.~\eqref{eq:unitarity-22}--\eqref{eq:unitarity-33} we employ the generalized matrix notation. Here, only the $\2\to\2$ case is considered. Taking imaginary part of Eq.~\eqref{eq:formal-solution-22-2} one arrives to
    \beq
    \label{eq:proof-22-1}
    \im \Ac_{22} &=& \im [\one -\Hc_{22} i \rho_2]^{-1} \Hc_{22} \nln
    & & + [\one + \Hc_{22}^* i \rho_2]^{-1} \im \Hc_{22} \, .
    \eeq
The first term above can be expanded as
    \beq
    & & \im [\one - \Hc_{22} i \rho_2 ]^{-1} =  \nln
    & & [\one + \Hc_{22}^* i \rho_2 ]^{-1} \Hc_{22}^* \rho_2 [\one - \Hc_{22} i \rho_2 ]^{-1} 
    \nln
    & & +  [\one + \Hc_{22}^* i \rho_2 ]^{-1} \im \Hc_{22} i \rho_2 [\one - \Hc_{22} i \rho_2 ]^{-1} \, ,
    \eeq
which used in Eq.~\eqref{eq:proof-22-1} together with the formal solution for $\wAc_{22}$, given in Eq.~\eqref{eq:formal-solution-22-2}, gives,
    \beq
    \label{eq:proof-22-2}
    \im \Ac_{22} &=& \Ac_{22}^* \rho_2 \Ac_{22} \nln
    & & + [\one + \Hc_{22}^* i \rho_2 ]^{-1} \im \Hc_{22} (1 + i \rho_2 \Ac_{22}) \, .
    \eeq
Also, since $\Hc_{22} = \Bc_{22} + \Bc_{23} \Fc [\one - \Bc_{33} \Fc ]^{-1} \Bc_{32}$ and
    \beq
    \im \Hc_{22} &=& \Bc_{23} \im \Fc  [\one - \Bc_{33} \Fc ]^{-1} \Bc_{32} \nln
    & & + \Bc_{23} \Fc^* \im [\one - \Bc_{33} \Fc ]^{-1} \Bc_{32} \, ,
    \eeq
where
    \beq
    & & \im [\one - \Bc_{33} \Fc ]^{-1} =\nln
    & &  [\one - \Bc_{33}^*\Fc^* ]^{-1} \im \Bc_{33} \Fc  [\one - \Bc_{33} \Fc ]^{-1} \nln
    & & +  [\one - \Bc_{33}^* \Fc^* ]^{-1} \Bc_{33}^* \im \Fc  [\one - \Bc_{33} \Fc ]^{-1} \, ,
    \eeq
the second term of Eq.~\eqref{eq:proof-22-2} can be transformed accordingly,
\begin{widetext}
    \beq
    \label{eq:proof-22-3}
    \im \Ac_{22} &=& \Ac_{22}^* \rho_2 \Ac_{22} + \Big( [\one + \Hc_{22}^* i \rho_2]^{-1} \Bc_{23} \im \Fc  [\one - \Bc_{33} \Fc ]^{-1} \nln
    & & + [\one + \Hc_{22}^* i \rho_2]^{-1} \Bc_{23} \Fc^* [\one - \Bc_{33}^*\Fc^* ]^{-1} \im \Bc_{33} \Fc  [\one - \Bc_{33} \Fc ]^{-1}  \nln
    & & + [\one + \Hc_{22}^* i \rho_2]^{-1} \Bc_{23} \Fc^* [\one - \Bc_{33}^* \Fc^* ]^{-1} \Bc_{33}^* \im \Fc  [\one - \Bc_{33} \Fc ]^{-1} \Big) \times (\Bc_{32} + \Bc_{32} i \rho_2 \Ac_{22}) \, .
    \eeq
The last step is to match the terms in the above equation with those of Eq.~\eqref{eq:unitarity-22}. First, it is useful to notice that $\wAc_{32} = [\one - \Bc_{33} \Fc ]^{-1} (\Bc_{32} + \Bc_{32} i \rho_2 \Ac_{22})$. Next one can employ the unitarity relation for the isobar amplitude $\im \Fc = \Fc^* \rho \Fc$. Finally, the imaginary part of the $B$-matrix kernel is rewritten as $\im \Bc_{33} = \im \Gc = \Cc$. After these three steps, one obtains,
    \beq
    \label{eq:proof-22-4}
    \im \Ac_{22} &=& \Ac_{22}^* \rho_2 \Ac_{22} +  [\one + \Hc_{22}^* i \rho_2]^{-1} \Big( \Bc_{23} + \Bc_{23} \Fc^* [\one - \Bc_{33}^* \Fc^* ]^{-1} \Bc_{33}^* \Big) \im \Fc \wAc_{32} \nln
    & & + [\one + \Hc_{22}^* i \rho_2]^{-1} \Bc_{23} \Fc^* [\one - \Bc_{33}^*\Fc^* ]^{-1} \im \Bc_{33} \Fc \wAc_{32} \\
    &=& \Ac_{22}^* \rho_2 \Ac_{22} + \wAc_{23}^* \Fc^* \rho \Fc \wAc_{32} + \wAc_{32}^* \Fc^* \Cc \Fc \wAc_{32} \, ,
    \eeq
which agrees with the two-body unitarity relation in Eq.~\eqref{eq:unitarity-22} after the amputation procedure. The last term in the above expression was obtained from iterating once the geometric series $[\one - \Bc_{33}^*\Fc^* ]^{-1} = \one + [\one - \Bc_{33}^*\Fc^* ]^{-1} \Bc_{33}^* \Fc^*$, that is
    \beq
    & & [\one + \Hc_{22}^* i \rho_2]^{-1} \Bc_{23} \Fc^* [\one - \Bc_{33}^*\Fc^* ]^{-1} \im \Bc_{33} \Fc \wAc_{32} \nln
    & & = [\one + \Hc_{22}^* i \rho_2]^{-1} \left( \Bc_{23} + \Bc_{23} \Fc^*  [\one - \Bc_{33}^* \Fc^* ]^{-1} \Bc_{33}^* \right) \Fc^*  \im \Bc_{33} \Fc \wAc_{32} \nln
    & & = \wAc_{23}^* \Fc^* \im \Bc_{33} \Fc \wAc_{32} \, .
    \eeq
Analogous calculations performed on Eqs.~\eqref{eq:formal-solution-33}, \eqref{eq:formal-solution-32} and \eqref{eq:formal-solution-23-2}, following the procedure shown here and in Ref.~\citep{Jackura:2018}, show that the $B$-matrix parametrization satisfies unitarity above the three-body threshold also in the case of the other amputated amplitudes.
\end{widetext}

\section{Spectrum of the contact interaction model}
\label{app:D}

\begin{figure}[hb!]
    \centering
    \includegraphics[width=0.4\textwidth]{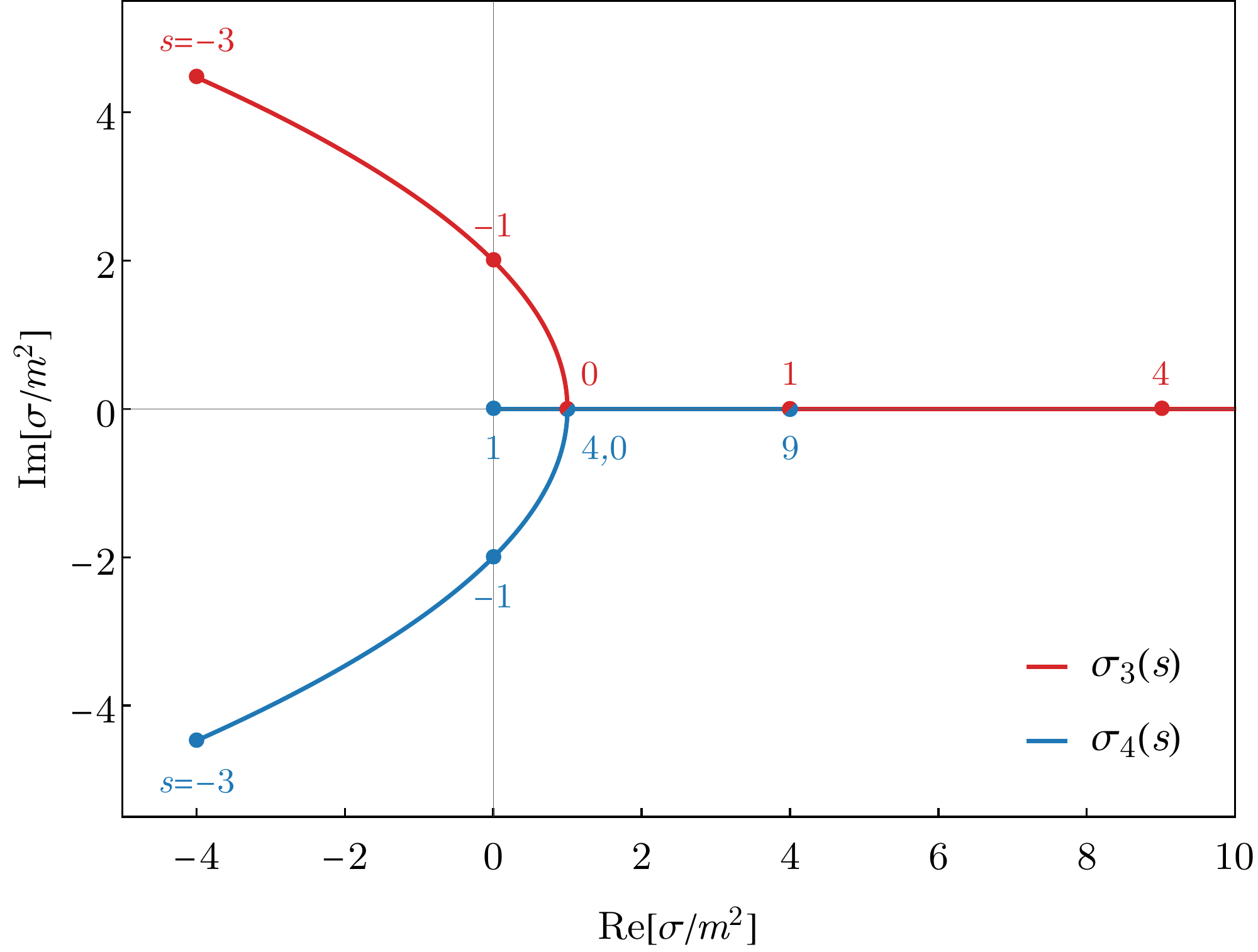}
    \caption{The trajectories of the movable singularities $\sigma_3$ and $\sigma_4$ as a functions of $s$. Here $s$ is decreasing from $s/m^2=9$ to $-3$. The $\sigma_3$ point travels on the real axis until it reaches $s/m^2=1$, where it becomes complex, with a positive imaginary part. The $\sigma_4$ moves on the real axis to the left until it reaches value $\sigma_4 = 0$ for $s/m^2=1$ and then it moves to the right, and becomes complex for $s=0$.  }
    \label{fig:mov_sin}
\end{figure}

\begin{figure*}[ht!]
\begin{center}
\subfigure[~$s/m^2=10$]
{
\includegraphics[width=0.4\textwidth]{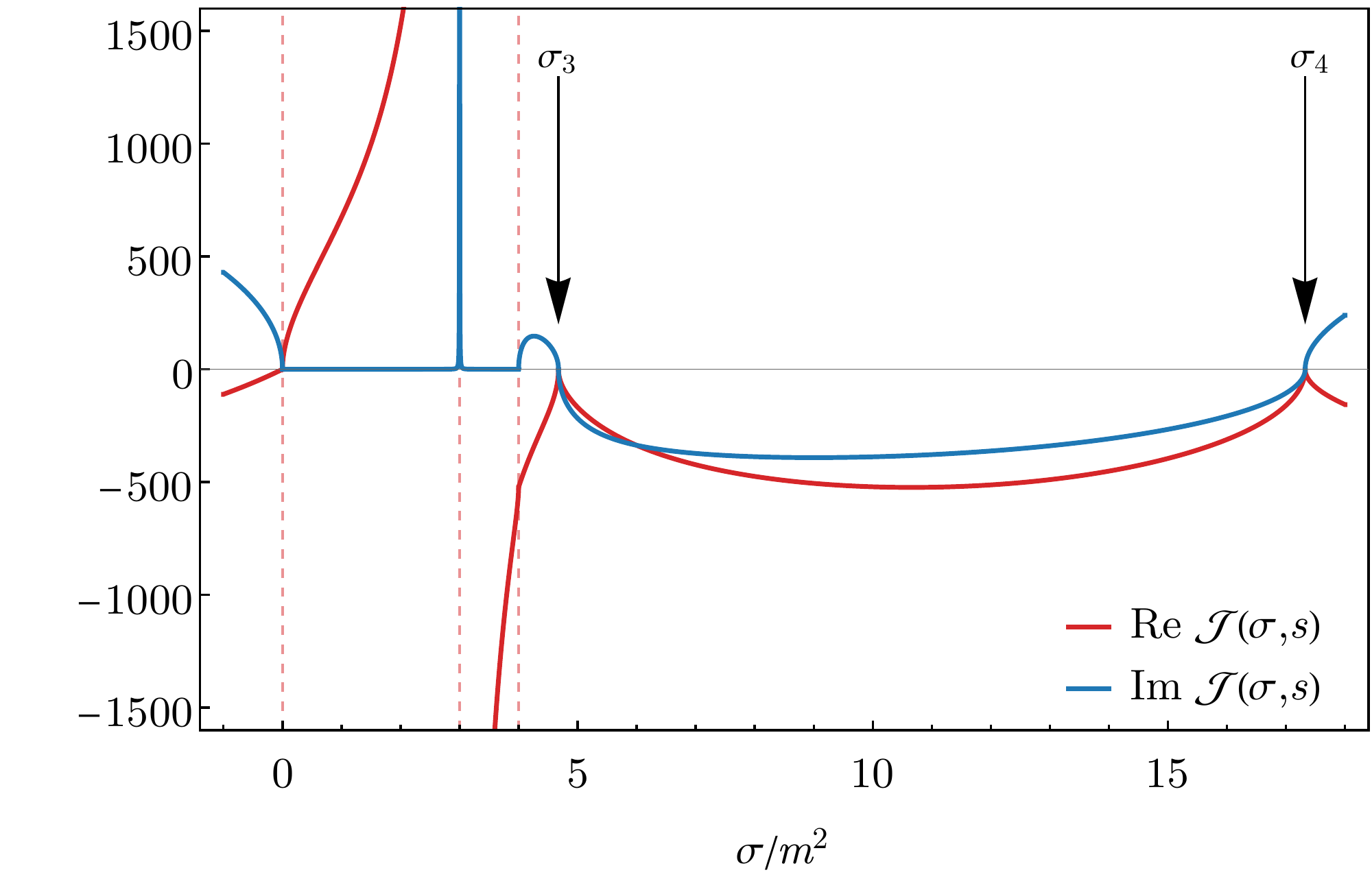}
}
\subfigure[~$s/m^2=9$]
{
\includegraphics[width=0.4\textwidth]{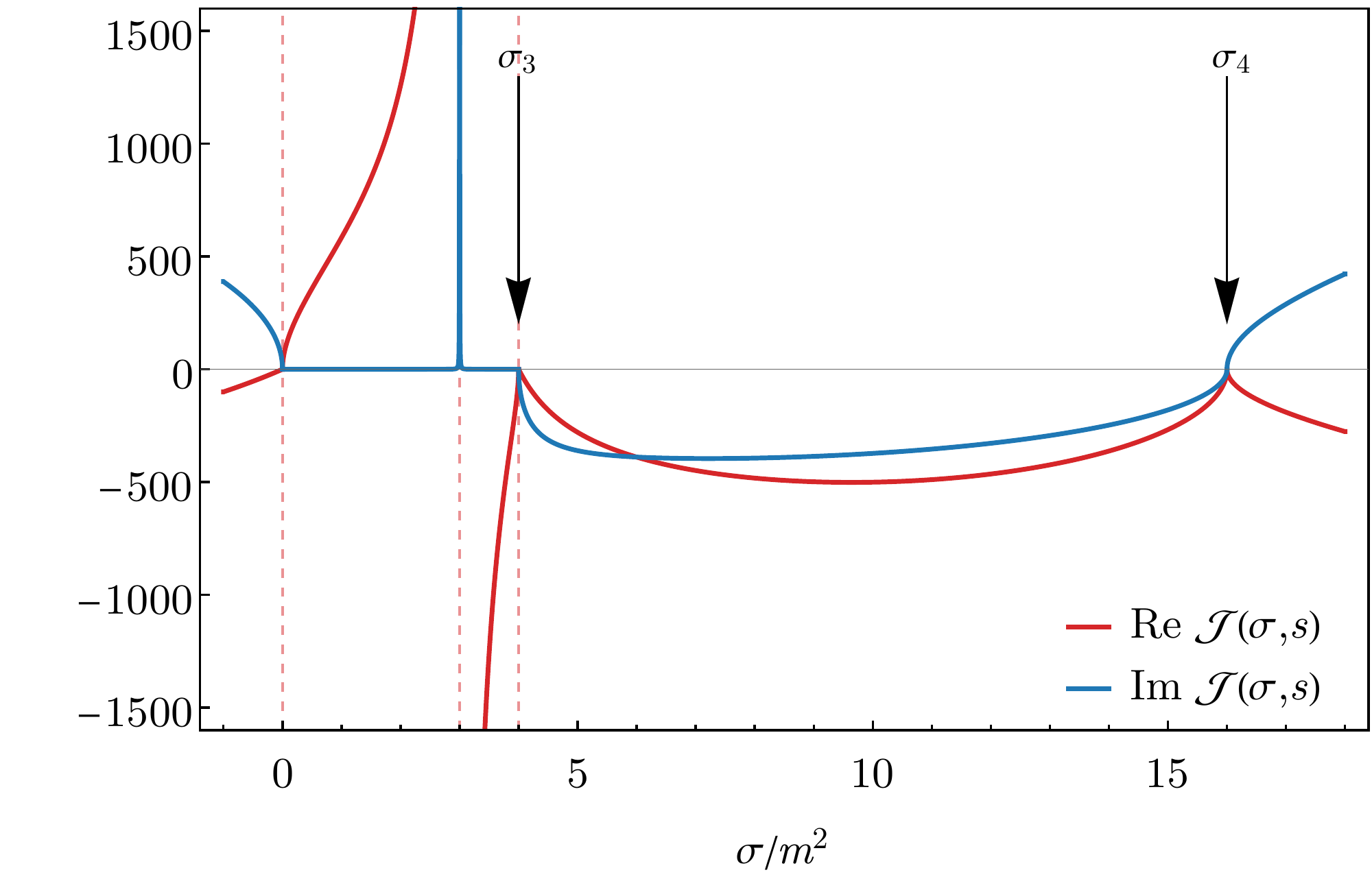}
}
\subfigure[~$s/m^2=8.35$]
{
\includegraphics[width=0.4\textwidth]{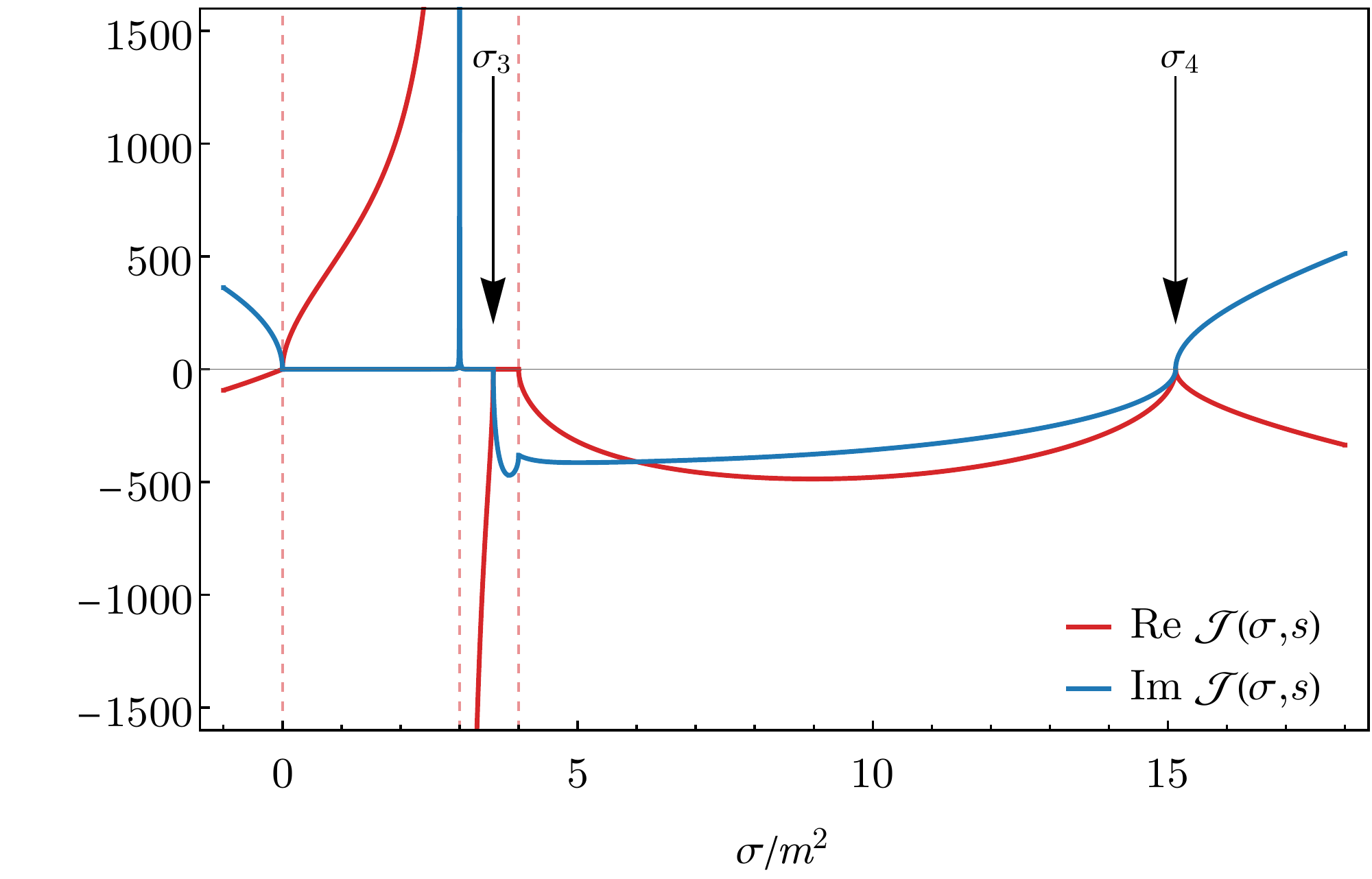}
}
\subfigure[~$s = s_{\text{th},2} \approx 7.456 m^2$. ]
{
\includegraphics[width=0.4\textwidth]{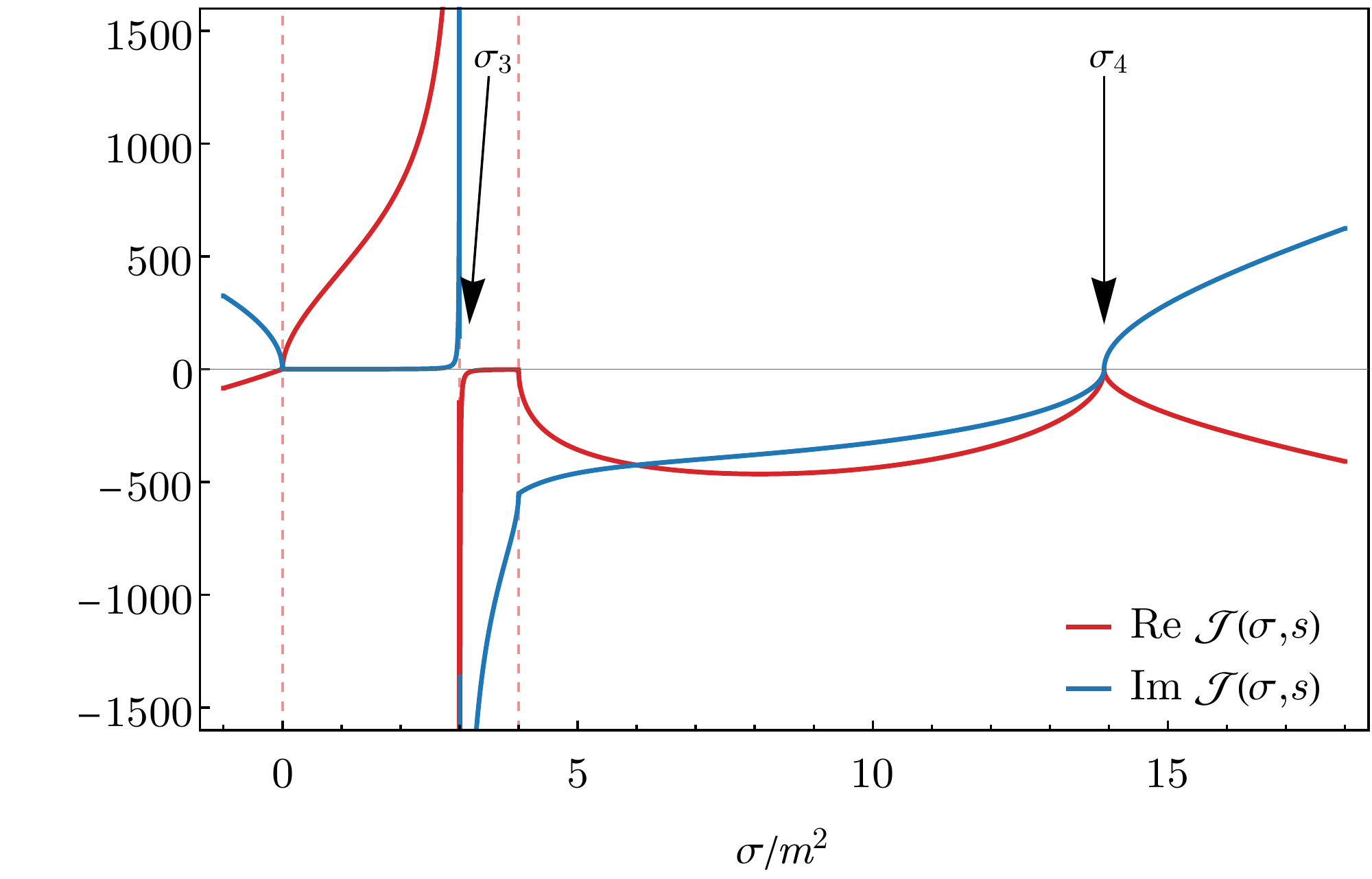} 
}
\subfigure[~$s/m^2 = 6$ ]
{
\includegraphics[width=0.4\textwidth]{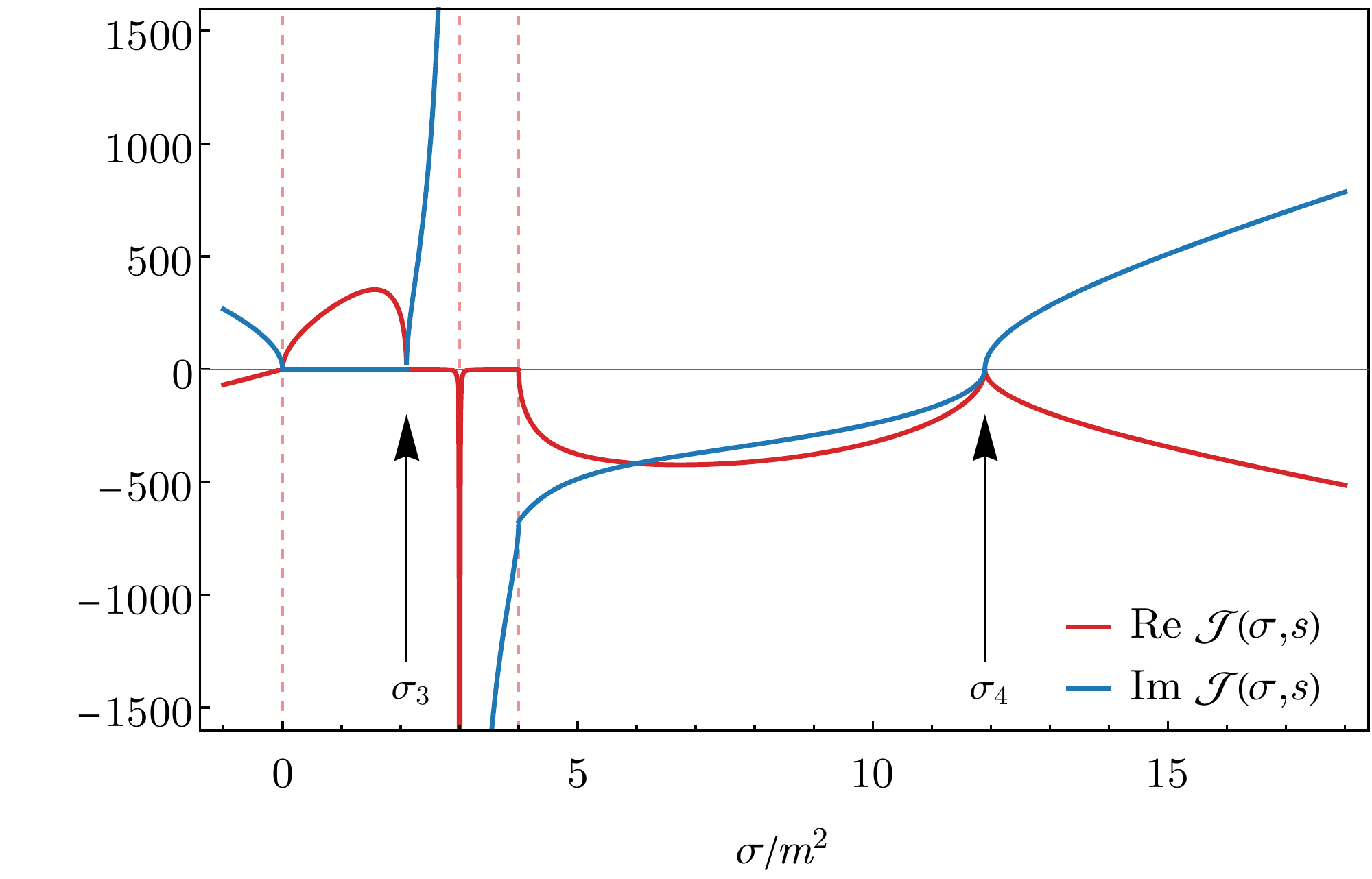}
}
\subfigure[~$s/m^2 = 1$. ]
{
\includegraphics[width=0.4\textwidth]{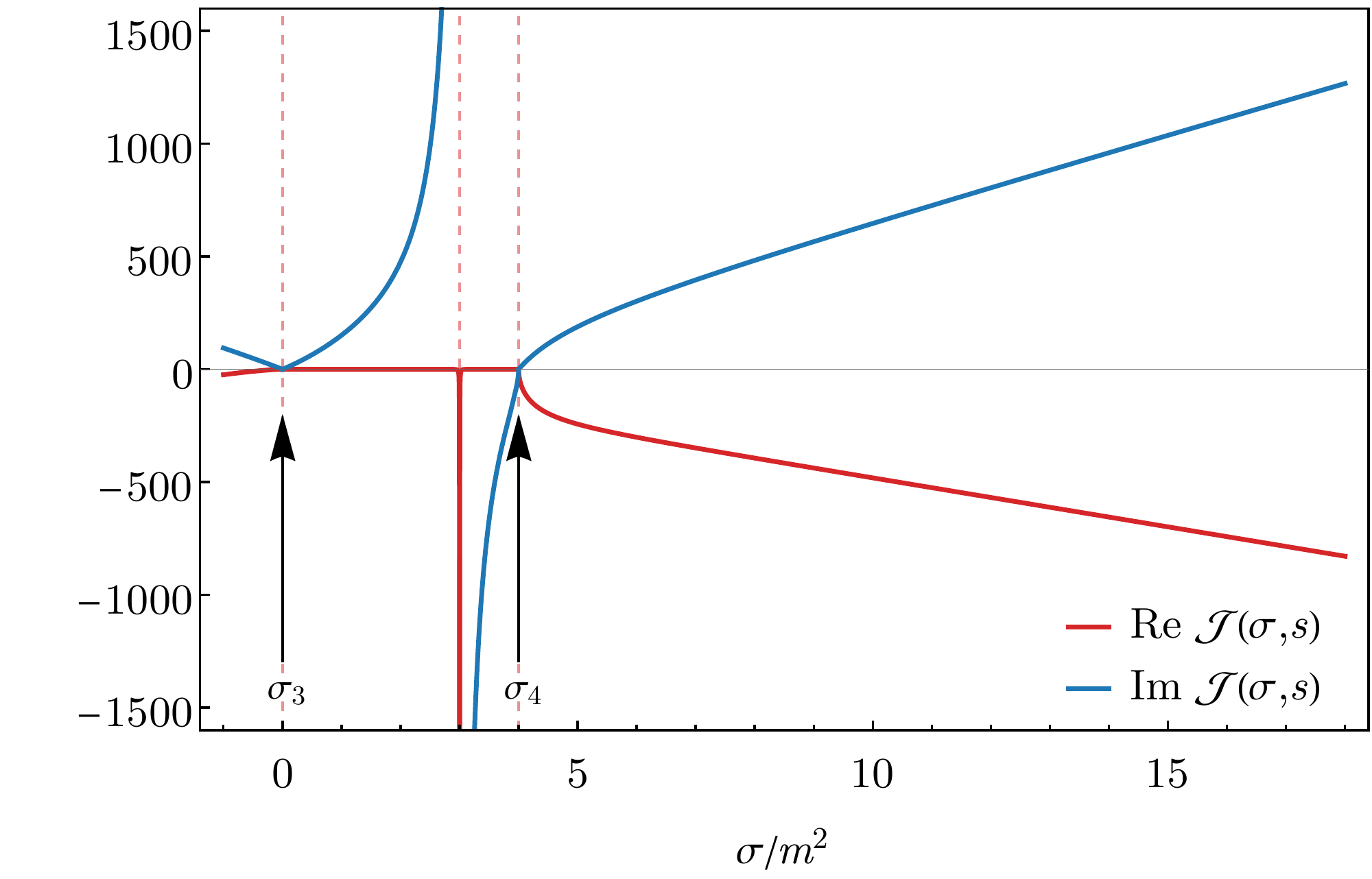}
}
\subfigure[~$s/m^2=0.1$. ]
{
\includegraphics[width=0.4\textwidth]{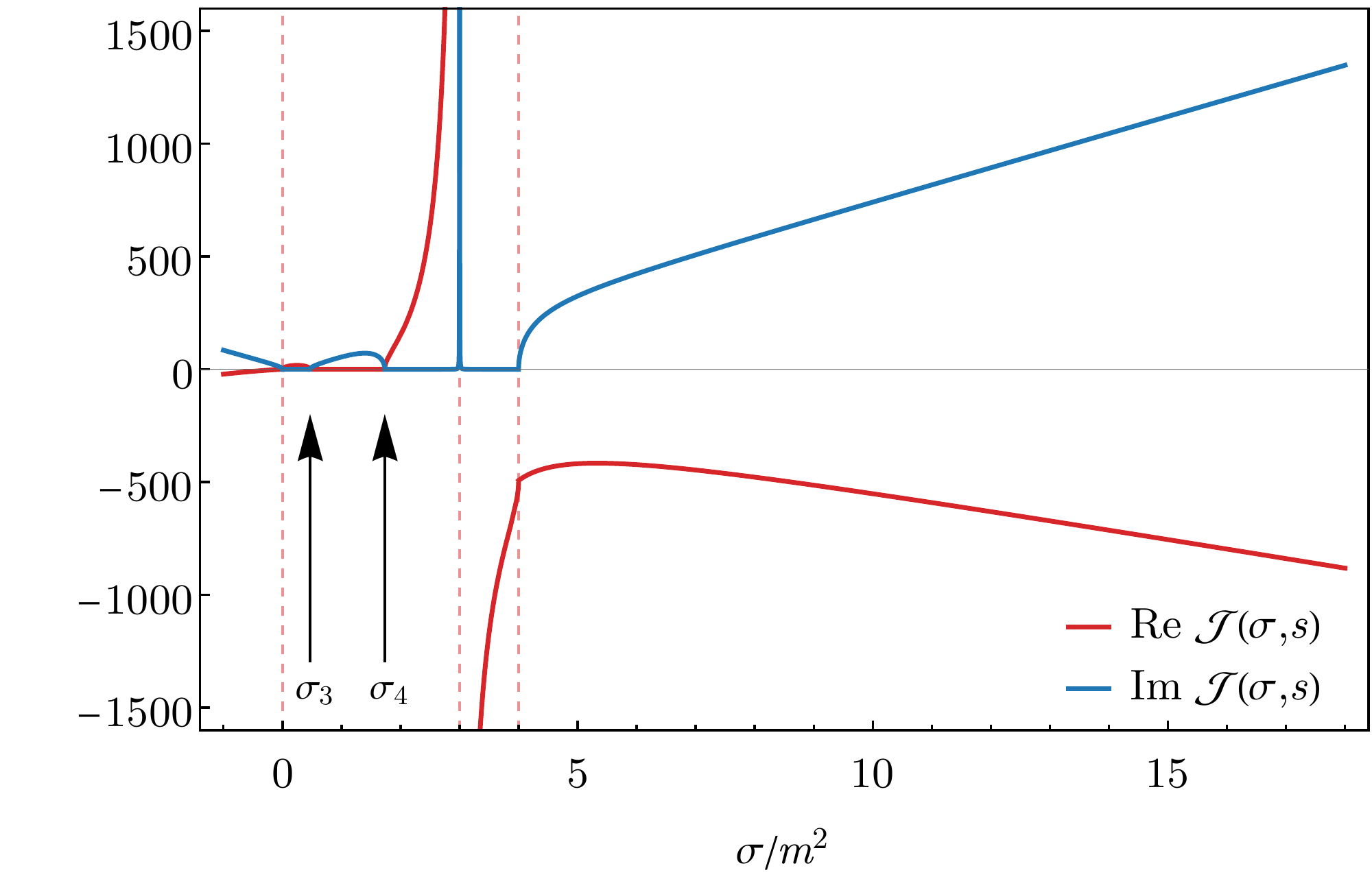}
}
\subfigure[~$s/m^2 = 0$. ]
{
\includegraphics[width=0.4\textwidth]{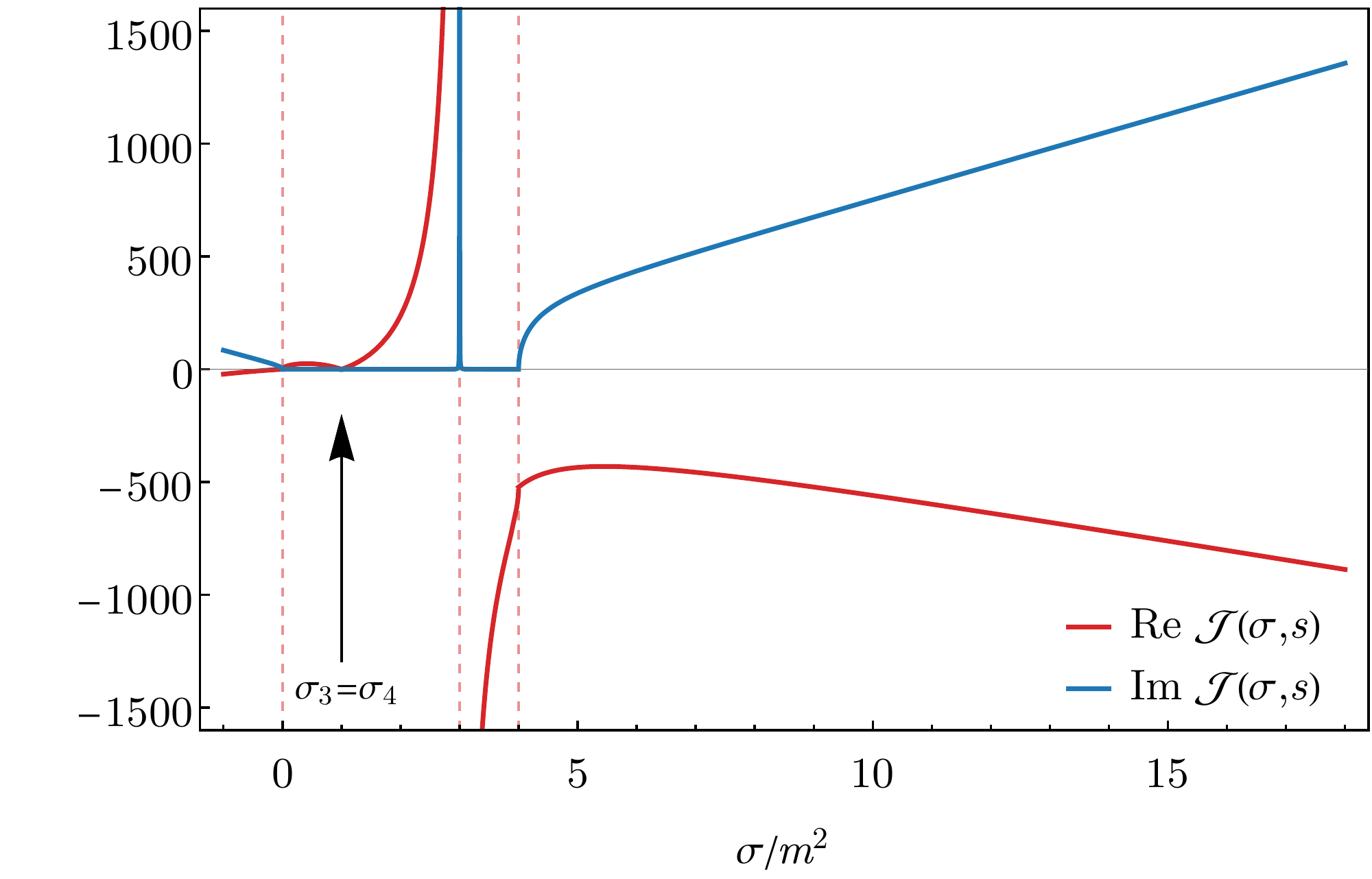}
}
\end{center}
\caption{The integrand $\Jc(\sigma, s)$ of Eq.~\eqref{eq:integrand} as a function of $\sigma$ for fixed, real values of $s$, decreasing from $s/m^2 = 10$ to $s/m^2=0$. The bound state pole occurs at $M^2 = 3 m^2$. The $s$-dependent branch points $\sigma_{3,4}$ are specified by arrows, while fixed points $\sigma_0/m^2=0, \sigma_b/m^2 = 3, \sigma_2/m^2=4$ are marked with dashed vertical lines. The plot was obtained for $\epsilon/m^2=10^{-4}$. \\~}
\label{fig:functJflow}
\end{figure*}

In this appendix, two figures which illustrate the discussion of the integrand $\Jc(\sigma,s)$ in Sec.~\ref{sec:contact} are included. On Fig.~\ref{fig:mov_sin} the motion of the singularities $\sigma_{3,4}(s)$ with the changing invariant mass squared $s$ is presented. Figure~\ref{fig:functJflow} shows the changes of the real and imaginary parts of $\Jc(\sigma,s)$ for $s$ decreasing from $s/m^2=10$ to $s/m^2=0$.

\onecolumngrid

\twocolumngrid
\clearpage

\bibliographystyle{apsrev4-1}
\bibliography{main}

\end{document}